\documentclass{emulateapj}

\def\lappr{\hbox{$_<\atop{^\sim}$}}

\def\Sec{${}^{\prime\prime}$\llap{.}}

\def\etal{{\it et~al.\/}}

\def\kpc-1{{kpc$^{-1}$}}
\def\Mpc-1{{Mpc$^{-1}$}}
\def\s-1{{sec$^{-1}$}}
\def\pdeg2{{deg$^{-2}$}}

\def\h0{{H$_0$}}
\def\q0{{$q_0$}}

\def\etal{{et al.}}

\def\ltsima{$\scriptscriptstyle \; \buildrel < \over \sim \;$}
\def\simlt{\lower.3ex\hbox{\ltsima}}

\def\gtsima{$\scriptscriptstyle \; \buildrel > \over \sim \;$}
\def\simgt{\lower.3ex\hbox{\gtsima}}

\def\about{\raise.3ex\hbox{$\scriptscriptstyle \sim $}}

\def\Sec{\hbox{${}^{\prime\prime}$\llap{.}}}

\def\sqr#1#2{{\vcenter{\vbox{\hrule height.#2pt
        \hbox{\vrule width.#2pt height#1pt \kern#1pt
        \vrule width.#2pt}
        \hrule height.#2pt}}}}
\def\square{{\mathchoice\sqr62\sqr62\sqr{4.2}1\sqr{3}1}\,}

\shortauthors{Kelson \etal}
\shorttitle{Line Strengths in E/S0 Galaxies at $z$=0.33}
\slugcomment{Accepted for Publication in ApJ}

\begin{document}

\title{
Line Strengths in Early-Type Cluster Galaxies at $z$=0.33:\break
Implications for $\alpha$/Fe, Nitrogen and the Histories of E/S0s
\altaffilmark{1}} 

\author{Daniel D.~Kelson\altaffilmark{2},
Garth D.~Illingworth\altaffilmark{3},
Marijn Franx\altaffilmark{4}, and
P.~G.~van Dokkum\altaffilmark{5}}

\altaffiltext{1}{Based on observations obtained at the W. M. Keck
Observatory, which is operated jointly by the California Institute of
Technology and the University of California.}

\altaffiltext{2}{The Observatories of the Carnegie Institution of
Washington, 813 Santa Barbara St, Pasadena, CA 91101; kelson/at/ociw.edu}

\altaffiltext{3}{UCO/Lick Observatories, University of
California, Santa Cruz, CA 95065; gdi/at/ucolick.org}

\altaffiltext{4}{Leiden Observatory, P.O. Box 9513, 2300 RA, Leiden, The
Netherlands; franx/at/strw.leidenuniv.nl}

\altaffiltext{5}{Yale University, New Haven, CT 06520;
dokkum/at/astro.yale.edu}

\begin{abstract}

In this paper we analyze previously published spectra with high
signal-to-noise ratios of E and S0 galaxies in the rich cluster
CL1358+62 at $z=0.33$, and introduce techniques for fitting stellar
population models to the data. These data and methods will be used
further in a larger study of the evolution of absorption line strengths
in intermediate redshift clusters. Here we focus on the 19 elliptical
and lenticular galaxies with an homogeneous set of eight blue Lick/IDS
indices. These early-type galaxies follow very narrow line strength-line
width relations using Balmer and metal lines, indicating a high degree
of uniformity in their formation and enrichment histories. We explore
these histories using recently published, six-parameter stellar
population models \citep{thomas,thomas2}, and describe a novel approach
for fitting these models {\it differentially\/}, such that the largest
sources of systematic error are avoided. The results of the model
fitting are accurate {\it relative\/} measures of the stellar population
parameters, with typical formal errors of $\simlt 0.1$ dex. The best-fit
models yield a mean $\chi^2$ of 1.2 per degree of freedom, indicating
that the models provide good descriptions of the underlying stellar
populations. We find: (1) no significant differences between the
best-fit stellar population parameters of Es and S0s at fixed velocity
dispersion; (2) the stellar populations of the Es and S0s are uniformly
old, consistent with results previously published using the
fundamental plane; (3) a significant correlation of [Z/H] with galaxy
velocity dispersion, in a manner consistent with the observed $B-V$
colors of the galaxies, and indicating that dust is not a significant
contributor to the colors of early-type galaxies; (4) a possible, modest
anti-correlation of [$\alpha$/Fe] with velocity dispersion, with $<
2\sigma$ significance, and discrepant with the correlation inferred from
data on nearby galaxies at the $<3\sigma$ level; and (5) a significant
anti-correlation of [$\alpha$/N] with galaxy velocity dispersion, which
we interpret as a correlation of nitrogen enhancement with mean
metallicity. Neither [$\alpha$/C], nor [$\alpha$/Ca] shows significant
variation. While the differences between our conclusions and the current
view of stellar populations may point to serious deficiencies, our
deduced correlation of mean metallicity with velocity dispersion does
reproduce the observed colors of the galaxies, as well as the slope of
the local Mg-$\sigma$ relation. Our tests indicate that the inferred
population trends do describe real galaxies quite well, and matching our
results with published data on nearby galaxies, we infer that the
discrepancy stems largely from the historical treatment of broadening
corrections to the narrow indices. The data also strongly indicate that
secondary nitrogen is an important component in the chemistry of
elliptical and lenticular galaxies. Taken together, these results reduce
early-type galaxies in clusters to a family with one-parameter, velocity
dispersion, greatly simplifying scenarios for their formation and
evolution. More specifically, our data conclusively show that cluster
S0s did not form their stars at significantly later epochs than cluster
ellipticals of the same mass, and the presence of secondary nitrogen
indicates that both Es and S0s formed from self-enriching progenitors,
presumably with extended star-formation histories. 

\end{abstract}

\keywords{
galaxies: clusters: individual (CL1358+62),
galaxies: stellar content,
galaxies: elliptical and lenticular, galaxies: evolution}


\section{Introduction}

Ever since the discovery that early-type galaxies follow scaling
relations \citep{mm1957}, it was clear they had profound implications
for cosmology as well as for galactic structure, formation, and
evolution \cite[e.g.][and subsequent
literature]{minkowski1962,terlevich}. Perhaps the most well-studied of
these scaling relations is the color-magnitude relation \citep{baum},
and its interpretation as a correlation of metal abundance with galaxy
luminosity \citep{rood1969} has survived to the present-day. Over the
past several decades, numerous other scaling relations have been
discovered and interpreted, such as $L$-$\sigma$ \citep{fj76},
$r_e$-$I_e$-$\sigma$ \citep[the fundamental plane;][]{dress7s,dd87},
Mg-$\sigma$ \citep{terlevich}, and other line strength-line width
relations, such as H$\delta_A$- or H$\gamma_A$-$\sigma$
\citep[e.g.][]{kuntschner2000,kelson01}.

Several important breakthroughs have helped to constrain the nature of
early-type galaxy scaling relations. The clearest results have come from
explicit tests for changes in the ages of stellar population with
redshift along the sequence of early-types. By measuring the slopes of
the color-magnitude relations in clusters to redshifts of unity
\cite{stanford} concluded that the relation originates largely from
systematic variations in metallicity with galaxy mass \citep[also
see][]{blakeslee2006}. Likewise, the fundamental plane of early-type
galaxies has a slope that does not appear to evolve significantly with
redshift to $z=0.33$ \citep{kelsonc}, further evidence that cluster
E/S0s are a family well-described by uniform ages and a mass-metallicity
relation. At redshifts of $z=0.8-0.9$ there are hints that the slope of
the fundamental plane may evolve modestly \citep{jorgensen2006}, though
this result appears to be inconsistent with the slope of the
color-magnitude relation as measured by \cite{blakeslee2006} and it
remains to be seen how these data will be reconciled.

There have been many efforts to observe the high-redshift line
strength-line width relations
\citep{ziegler,kelson01,jorgensen0152,barr2005,moran2005}. However,
significant constraints have not been forthcoming because of the
difficulty of obtaining sufficient signal-to-noise ratios.

The situation is much better at low redshifts. Detailed analysis of line
strengths in low-redshift E/S0s have been very revealing since
\cite{jesus} (and others) broke the age-metallicity degeneracy
\citep{worthey}. Using line strengths many authors have shown that
massive galaxies in clusters are uniformly old
\citep[e.g.][]{jorgensen1997,jorgensen1999,trager2000a,kuntschner2000},
but some ambiguities do remain. For example it has long been understood
that the Mg$b$-$\sigma$ \citep{terlevich} and Fe-$\sigma$ \citep{jesus}
relations are not consistent with the hypothesis that they both arise
solely from a correlation between the mean metallicity of the stellar
populations and galaxy mass. The discrepancy has been interpreted using
stellar populations models in which ``$\alpha$-enhancement, or
[$\alpha$/Fe], is allowed to vary systematically with velocity
dispersion \citep[see, e.g.][for details and many important
references]{jorgensen1999,trager2000a,trager2000,worthey2003,thomas2005}.
While \cite{trager2000a} reminded readers that this is a misnomer,
because the $\alpha$ elements are not enhanced --- Fe is simply
under-abundant. We will, however, continue to use the term
``$\alpha$-enhancement'' to be consistent with past literature on the
subject. After several decades of analysis, improved techniques of
observation, and despite the immense amount of progress in stellar
population models
\citep{faber1972,worthey,trager2000a,schiavon1,thomas}, the modeling of
absorption lines remains ambiguous and contradictory \citep[e.g.,][and
others]{worthey,trager2000,worthey2003,schiavon3,thomas-ca}. Some have
even questioned the ``$\alpha$-enhancements'' altogether
\citep[]{proctor2004a}.

Over the long-term we hope that the physics of stellar atmospheres will
be understood with sufficient detail to allow for the creation of
synthetic stellar spectra for stars, over the full range of stellar
masses and phases of stellar evolution, that comprise the observed
spectra of galaxies. Presently, however, numerous uncertainties in the
formation of molecular lines, and incomplete line lists prevent one from
generating completely synthetic spectra for stellar populations. As a
result the features in the spectra of stars and galaxies cannot be fully
modeled. Despite major progress in generating high resolution spectral
energy distributions of simple stellar populations \citep{vaz2a,bc2003},
we cannot fully solve for the age(s) and elemental abundances
of populations in a galaxy by directly fitting synthetic spectra.

\begin{figure*}
\centerline{\epsscale{0.7} \plotone{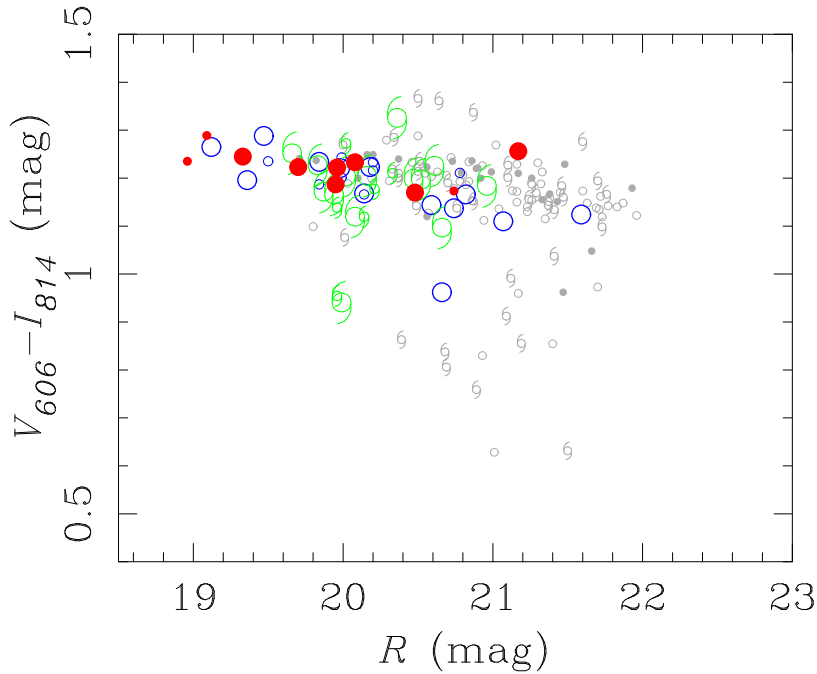}}
\caption{Color-magnitude diagram of the cluster in $R$ and
$V_{606}-I_{814}$, taken from \cite{kelsonb}. All 194 confirmed cluster
members in the HST imaging are shown. The red, blue, and green symbols
represent those galaxies observed by \cite{kelsonb} for use in their
study of the fundamental plane. Ellipticals are shown using red, filled
circles. The S0 galaxies are shown using blue open circles. Spiral
galaxies are shown in green. The light gray points were not observed in
the high-resolution study of \cite{kelsonb}. The red, blue, and green
points are shown using two sizes. The larger ones mark those galaxies in
the homogeneous sample (for which all eight of the indices used in the
modeling are available; see \S \ref{sec:fitting}). The E and S0 (red and
blue) galaxies in the figure are discussed in this paper while the full
fundamental plane sample of CL1358+62 will be discussed in a subsequent
paper.
\label{fig:selection}}
\end{figure*}

Without the ability to directly model the spectra of galaxies, one must
measure and model spectral indices. One widely used system is the
Lick/IDS system, created by \cite{burstein84} \citep[and subsequently
revised by][]{trager}. To this day, the measurement and modeling of
these line strengths remain the most useful means of assessing the bulk
properties of the stellar populations in passively evolving galaxies.

Over the last several years several groups have expanded our ability to
model these indices with more parameters than age, metallicity, and
$\alpha$-enhancement. The models of \cite{trager2000a},\cite{thomas},
and \cite{schiavon2005} allow one to probe specific elemental
abundances, such as nitrogen, carbon, and calcium. Such models only make
predictions for passively evolving stellar populations. Fortunately
models of galaxy evolution involving star-formation are not needed in
our study of massive cluster galaxies through intermediate redshifts,
because such galaxies have been shown to be passively evolving
\citep[e.g.,][]{kelson97,kelsonc,wuyts}. At higher redshifts, on-going
star-formation may become increasingly important \citep[e.g.][]{juneau},
and more sophisticated models may then be required.

In our survey of galaxies in rich clusters at intermediate redshifts,
the sample of galaxies with the highest signal-to-noise spectroscopy
over the widest range of galaxy luminosities is that from our
fundamental plane survey of CL1358+62 at $z=0.33$. Because of the depth
and quality of that sample, we adopt it as the reference sample to which
the other clusters in our survey will subsequently be compared
\citep{evolpaper}.

Here we study the absorption lines of the E/S0 galaxies in that sample,
using the models of \cite{thomas} and \cite{thomas2} to explore not only
ages and metallicities, but additional stellar population parameters,
namely the relative abundances of nitrogen, carbon, and calcium, along
with the mean $\alpha$-enhancement. Because we are keenly interested in
potential variations in the abundance ratios, and because the
state-of-the art high-resolution models \citep{bc2003} do not reproduce,
in detail, the strengths of many features in the spectra of real
galaxies \citep[e.g. CN, Ca4227, Mg, H$\delta$,
H$\gamma$;][]{gallazzi2005}, we prefer to derive stellar population
parameters from an analysis of absorption line indices. The
\cite{bc2003} SEDs are employed for other purposes and these are
discussed below in the context of deriving broadening corrections to our
data.

Many of the details regarding the processing of the data, and subsequent
corrections, are discussed elsewhere \citep{kelsonb,datapaper}. However
several of the key points are discussed below in \S \ref{sec:data}
because of their crucial role in the analysis. We then describe our
comparison of the absorption line strengths of the E/S0 galaxies to the
models of \cite{thomas} and \cite{thomas2}, in which we perform
non-linear least-squares fits to each galaxy's set of line strengths
\cite[see, also][]{proctor2004b}, {\it but only for relative differences
in the stellar population parameters\/}. This allows us to explicitly
avoid potentially large systematic uncertainties in the direct
comparison of the data and models. This methodology is used in the rest
of the survey and so is described in some detail. The resulting relative
ages and patterns of fitted chemical abundances are described in \S
\ref{sec:fits} and then discussed further in \S \ref{sec:summary}. Our
conclusions are summarized in \S \ref{sec:conclusions}. While our
findings do not depend on the cosmology, we use the cosmological
parameters $H0=72$ km/s/Mpc, $\Omega_M=0.27$, and $\Omega_\Lambda =
0.73$, when such parameters are required (e.g. for corrections to the
line strengths for any variation in the metric sizes of the apertures
from which the spectra were obtained).


\section{The Data}
\label{sec:data}

\begin{figure*}
\centerline{\epsscale{0.9} \plotone{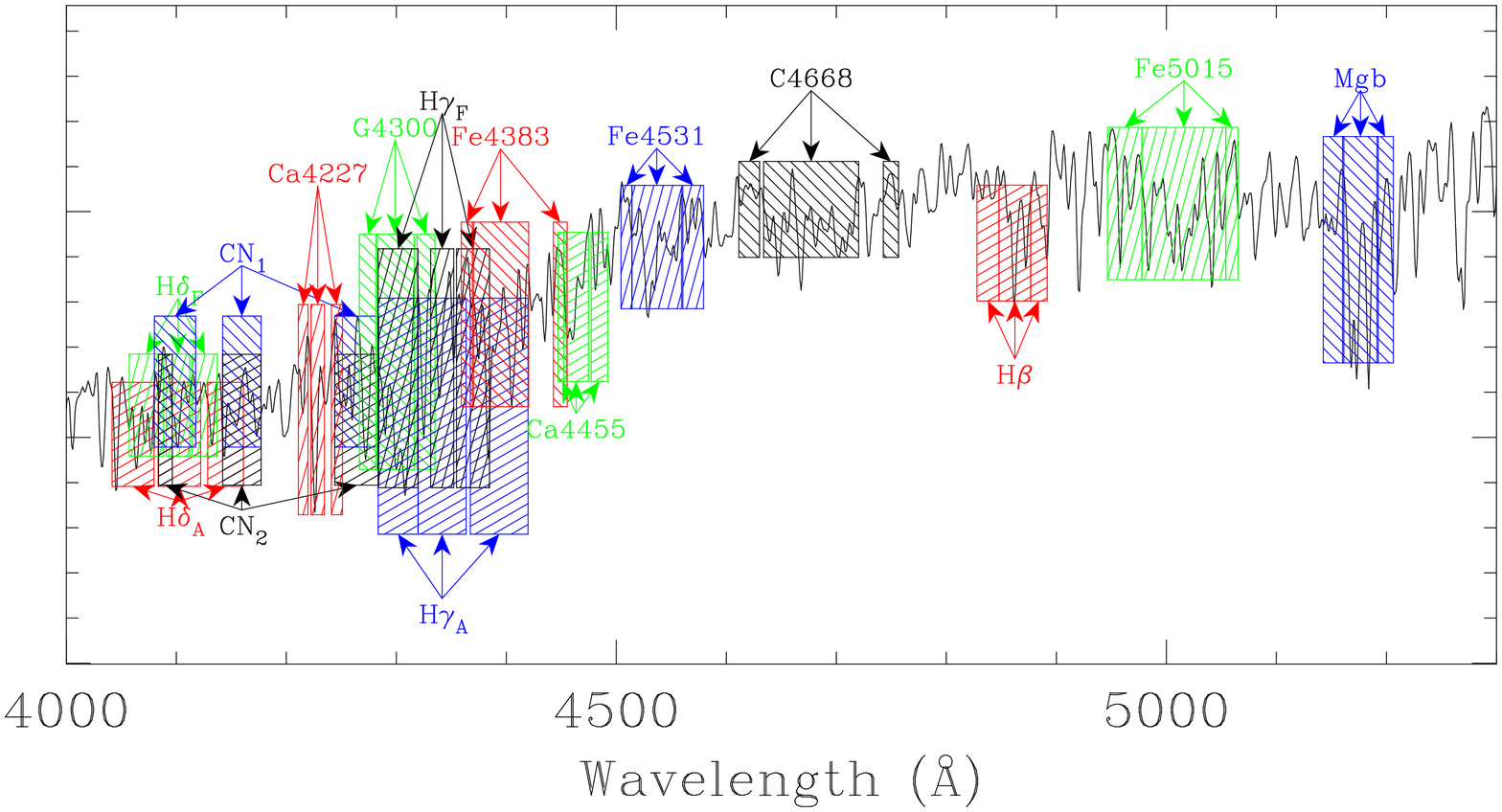}}
\caption{Overlayed on a model spectrum of a 6 Gyr-old population
\citep{bc2003} is the set of Lick/IDS bandpasses used in this paper
\citep{wortheys,trager}
\label{fig:diagram}}
\end{figure*}

As stated earlier, the galaxies in CL1358+62 were originally targeted as
part of a survey to study the fundamental plane in intermediate redshift
clusters \citep[published by][]{kelson97,vdfp83,kelsonc,wuyts}. The
selection of the CL1358+62 spectroscopic sample, and data processing,
were fully described in \cite{kelsonb}, and we summarize it here.

The field of CL1358+62 was targeted for extensive imaging with the HST
WFPC2. A two-color mosaic of the cluster CL1358+62 covering $64\,
\square\, '$ was constructed using 12 pointings. Of the
spectroscopically confirmed members \citep{fish}, 194 fall within the
field of view of the WFPC2 imaging. From this large catalog, we selected
a sample for detailed study with the Low Resolution Imaging Spectrograph
\citep[LRIS;][]{okelris} at the W.M. Keck Observatory. The high
resolution of the HST imaging allowed us to derive structural parameters
with the accuracy needed for the fundamental plane
\citep{kelsona,kelsonc}.

The sample used in the fundamental plane analysis was randomly selected
from a catalog of known members within the field of the HST mosaic down
to $R < 21$ mag. Because the selection did not rely on morphological
information, the galaxies span a range of morphologies, from ellipticals
to spirals. In Figure \ref{fig:selection}, taken from \cite{kelsonb}, we
show the color-magnitude diagram for the spectroscopically confirmed
cluster members within the HST mosaic. At the redshift of the cluster,
$L^*$ corresponds to $R\sim 20.1$ mag \citep[adopting the $M/L_V$
evolution as measured by][]{kelsonc}. There is clearly a tight
color-magnitude relation in the cluster, fully analyzed in
\cite{vdcm33}. The entire catalog of spectroscopically confirmed cluster
members within the HST mosaic is shown by the circles, and those
selected for fundamental plane analysis are shown by the filled circles.
As can be seen in the figure, the sample fairly represents the cluster
population, minus the faint (and very blue) objects.

In \cite{fish} there are 108 members brighter than $R=21$ mag within the
HST mosaic. Our high-resolution spectroscopic sample contains 52 of
them. Three cluster galaxies fainter than the magnitude limit were
added, bringing the total number of the sample to 55. The spectra were
shown in \cite{kelsonb}, and the signal-to-noise ratios ranged from
about 15 per \AA\ to 90 per \AA. The 900/mm grating provided a typical
resolution of $\sigma_{inst}\approx 60$ km/s. Details of the reductions
to one-dimensional spectra were also given in \cite{kelsonb}.

Below we describe the methods for measuring the absorption line
strengths and velocity dispersions. The algorithm for deriving the
latter is necessary because we employ portions of it in determining the
corrections for Doppler and instrumental broadening, and for estimating
the formal uncertainties in the line strengths. Fuller discussions of
our measurements of the absorption line strengths, including the
procedures to estimate formal errors, the procedures to correct the line
strengths for the Doppler and instrumental broadenings, and the
corrections for the different metric aperture sizes are contained in
\cite{datapaper}.


\subsection{Absorption Line Strengths}

At this time the best understood and modeled diagnostics remain the
spectral indices of the Lick/IDS system
\citep{faber85,burstein84,burstein86,gorgas,wortheys}, defined as a set
of indices measuring the strengths of Balmer, metal, and molecular
features. We adopted the \cite{trager} definitions for the Lick/IDS
indices, with the additional definitions of \cite{wortheyhd} for the
indices of the higher order Balmer lines H$\delta$ and H$\gamma$. Each
index is comprised of a blue continuum bandpass, a central index
bandpass, and a red continuum bandpass.

The bandpasses of the blue indices are shown graphically in Figure
\ref{fig:diagram}. Note that the stellar absorption lines covered by
each index are heavily blended in galaxy spectra as a result of Doppler
broadening and smoothed further by the line-spread function of the
instrument. In order to compare line indices from one galaxy to another,
or from data taken with one instrumental set-up to data taken with
another, the indices must be corrected to identical levels of intrinsic
Doppler and instrumental broadenings, and these corrections are
discussed below. 

The absorption line strengths were measured using the following
prescription:

The straight line between the mean fluxes in the two continuum
bandpasses is defined as the local continuum, $C_\lambda$, within the
index bandpass. The mean flux within a bandpass is written as:
\begin{equation}
F_p = {1\over (\lambda_2-\lambda_1)}
\int_{\lambda_1}^{\lambda_2} F_\lambda d\lambda
\label{flux}
\end{equation}
Each index is then defined by the integral of the residual flux above
and/or below the local continuum within the index bandpass
\citep{worthey}. For convenience we define
\begin{equation}
I_p= {1\over (\lambda_2-\lambda_1)}
\int_{\lambda_1}^{\lambda_2}{F_\lambda\over C_\lambda}d\lambda
\label{integrand}
\end{equation}
The metal and Balmer line strengths have units of \AA\ and are commonly
referred to as pseudo-equivalent widths, while molecular features are
expressed as magnitudes:
\begin{eqnarray}
X_{\rm ew}  &=&  \bigl({\lambda_2-\lambda_1}\bigr)(1 - I_p) \\
X_{\rm mag}  &=&  -2.5 \log I_p
\label{mag}
\end{eqnarray}
The integrations are straightforward, with the contributions from
partial pixels computed using linear interpolation.


\subsection{Velocity Dispersions}

Many scaling relations of early-type galaxies use galaxy velocity
dispersion, $\sigma$, as the dependent variable
\citep[e.g.][]{fj76,terlevich,dd87,dress7s}. Reliable measurement of
velocity dispersion is also key to deriving reliable line strengths as
the velocity dispersions are required to compute the Doppler corrections
to the line strengths. The procedures used for deriving $\sigma$ are
also used in our method to estimate the formal errors in the line
strengths.

Our preferred method for measuring velocity dispersions is the
direct-fitting method \citep[described by][though also see
\citep{rix92}]{kelsonb}, It has several advantages over techniques that
operate in Fourier space \citep[e.g.][]{tonry,franx89}. Most
importantly, when measuring velocity dispersions of high-redshift
galaxies, non-uniform sources of noise become important, and these
affected pixels should not be accorded uniform weight. By fitting
spectra directly, one measures the velocity dispersion by shifting and
broadening a template spectrum until the result matches each galaxy
spectrum \citep{bbf1,rix92}. Mathematically, one has a galaxy spectrum,
$G$, and a template spectrum, $T$, and one searches for the
line-of-sight velocity distribution $B$, that is used to convolve $T$
such that $\chi^2$ is minimized:
\citep{kelsonb}:
\begin{equation}
\chi^2=\bigl|\bigl\{G-\bigl[P_M (B(\sigma,v)\circ T) +
\sum_{j=0}^{K}a_j H_j\bigr]\bigr\}\times W\bigr|^2.
\label{chi2}
\end{equation}

Following \cite{kelsonb} we parameterize the line-of-sight velocity
distribution as a Gaussian, $B(\sigma,v)$, in which $\sigma$ is the
velocity dispersion, or second moment of the velocity distribution and,
$v$ is the mean radial velocity, or first moment of the velocity
distribution. The fit also includes two additional components: $P_M$, a
low-order multiplicative polynomial that incorporates the difference
between the instrumental response function in $G$ and $T$, and $\sum a_j
H_j$ is an additive continuum function comprised of sines and cosines up
to wavenumber $K$ \citep{kelsonb}. These sines and cosine continuum
functions are equivalent to filtering low wavenumbers in the Fourier
domain \citep[see, e.g.,][]{tonry}.

The unique value of the direct-fitting method is contained in $W$, the
weighting spectrum. We opt, as in \cite{kelsonb}, to weight most pixels
by the inverse of the noise, estimated from photon statistics and
amplifier read noise. Just as in \cite{kelsonb}, there are exceptions:
(a) regions contaminated by galactic emission lines are given zero
weight; (b) strong Balmer absorption lines are also given zero weight;
and (c) pixels contaminated by poorly subtracted night-sky emission
lines are also given zero weight. Nearly all of the early-type galaxies
discussed in this paper have no (or very weak) emission. The BCG,
however, does have emission lines and therefore it is the only galaxy in
this paper for which this ``masking'' is important. The fitting was
performed between $\sim 4250$\AA\ and $\sim 5050$\AA, in the rest-frame,
employed 5th-order for the multiplicative polynomial, $P_M$, and sines
and cosines up to $K=6$. The results, as shown by \cite{kelsonb}, are
sensitive to these choices only at the level of a few percent.

The template spectra spanned a range of late-type stars, as was also
discussed previously in \cite{kelsonb}. Ideally one might wish to employ
spectra which represent more accurate models of the galaxy spectra.
However, the current state-of-the-art high-resolution model spectra
\citep{bc2003} have lower resolution ($\sim 3$\AA) than our instrumental
setup ($\sim 1.7$\AA). Therefore these model spectra are not suitable
for providing accurate measurements of velocity dispersion in our
lower-mass galaxies. If one derives velocity dispersions using the
\cite{bc2003} SEDs, the one must add, in quadrature, the difference in
resolution to the result. Doing so, we find $<1\%$ systematic difference
between those sigmas and the ones we obtained using the \cite{kelsonb}
stellar templates, with a standard deviation of 2\%. Furthermore, there
is no statistically significant trend with velocity dispersion for
galaxies with $\sigma>100$ km/s.


\subsection{Absorption Line Strength Errors}
\label{sec:errors}

Accurate estimates of the uncertainties in absorption line strengths
require one to determine the variances within each index bandpass
\citep[e.g.,][]{jesus,cardiel}. One can adopt the expected noise due to
photon statistics and amplifier read noise, but sources of noise become
increasingly localized for high redshift galaxy spectra, with the bright
OH lines and, potentially, fringing dominating the errors in flux.
Furthermore, the noise can become correlated in adjacent pixels after
the data are rebinned onto a simple wavelength scale \citep{cardiel}.

We developed an approach that provides for a more robust determination
of the line strengths and their errors, with less susceptibility to
systematic errors. This approach requires neither the additional
overhead of variance images nor any assessment of correlated noise in
adjacent pixels. Furthermore, our method does not rely on the assumption
that the ``true'' noise is well-represented by the theoretical
expectation from photon statistics and the electronics noise. Tests with
the data and simulations indicate that our methodology produces accurate
estimates of line strength errors. We discuss the method here.

Deriving the random errors in absorption line strengths requires
knowledge of the total variances in the continuum and index bandpasses.
Given an observed galaxy spectrum $G$, the noise in a given pixel is
simply $N=G-R$, where $R$ represents the underlying, noiseless spectrum
of the galaxy. While there is no {\it a priori\/} knowledge of $R$,
Equation \ref{chi2} allows us to write 
\begin{equation}
R \approx \widetilde R = P_M (B\circ T) + \sum_{j=0}^{K}a_j H_j
\label{model}
\end{equation}

Using the model stellar population spectra of \cite{bc2003}, we
generated the best-fit model SED for each galaxy. The model spectra were
used, instead of our template stars, because they better reproduce the
Balmer, metal, and molecular line strengths in $G$. One should note,
however, that in searching for the best-fit SED, no unique best-fit
age/metallicity pair could be found for any given galaxy: young,
metal-rich model SEDs fit as well as old, metal-poor model SEDs in all
cases, with complete degeneracy. Because individual line strengths are
subject to a similar degeneracy between age and metallicity
\citep[e.g.,][]{worthey}, attempts to match the detailed spectra are
subject to the mean correlation between age and metallicity because the
computation of $\chi^2$ utilizes all of the spectral features
simultaneously.

The approximation to $R$ provides an approximation to $N\approx
\widetilde N$.
\begin{equation}
\widetilde N = G - \bigl[P_M (B\circ T) + \sum_{j=0}^{K}a_j H_j\bigr]
\label{noise}
\end{equation}

With $\widetilde N$ the variance in the mean flux, $F_p$, is
\begin{equation}
V_{F_p} = {1\over {\lambda_2-\lambda_1}}
     \sum_{i}^{n} {\widetilde N_i}^2 \Delta\lambda_i
\label{fstderr}
\end{equation}
where the summation is over the pixels within the given bandpass and $n$
is the number of pixels within the bandpass.
For pixels of constant $\Delta\lambda$, this reduces to the familiar
\begin{equation}
V_{F_p} = {1\over n} \sum_{i}^{n} {\widetilde N_i}^2
\end{equation}
where the error in the mean flux is then
\begin{equation}
\varepsilon_{F_p} = \sqrt{{1\over n^2} \sum_{i}^{n} {\widetilde N_i}^2}
\end{equation}
Note that partial pixels are included at the ends of the bandpass by linearly
interpolating the fractional contribution of that pixel to the bandpass
and adding those to the summation. For clarity we did not include them
in the above equations.

\begin{figure*}
\centerline{\epsscale{0.95} \plotone{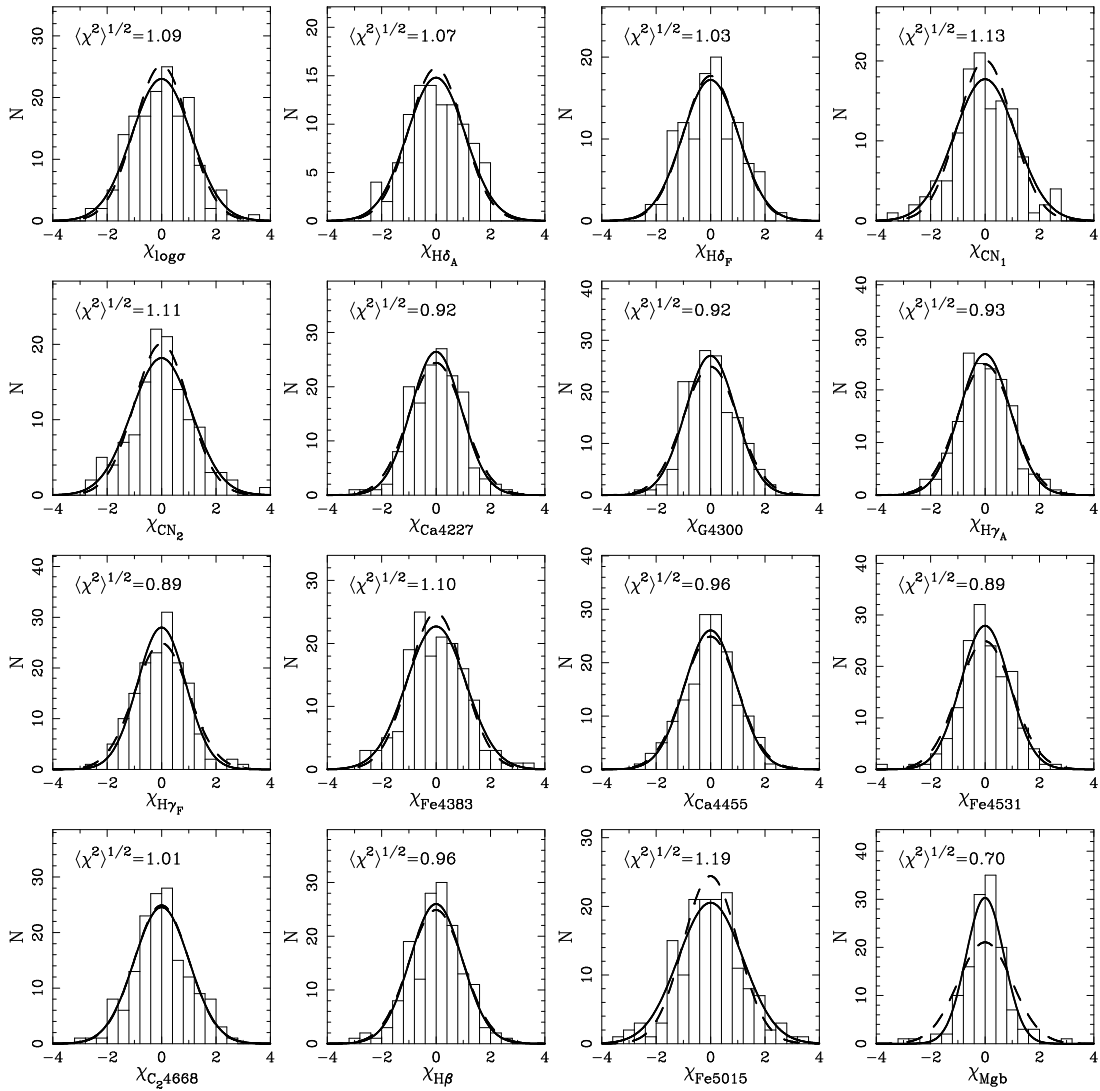}}
\caption{Histograms showing the variance in the measurements of the
indices from the individual exposures. Each galaxy's spectrum was
observed in three exposures and a comparison between the line strengths
in the individual exposures provides an independent test of our
estimates for the formal errors. The dashed curves show Gaussians with
standard deviations of unity to illustrate the expected distribution of
the data. The solid curves show Gaussians with standard deviations
determined empirically and listed within each panel. Excluding Mg$b$,
the mean standard deviation is 0.95, indicating that the formal errors
in our line strengths have been properly estimated. The Mg$b$ bandpass
suffers from telluric absorption and our correction to it is uncertain.
As a result the residual telluric absorption reduces the variance
between the individual observations.
\label{fig:histo}}
\end{figure*}

With accurate estimates of the errors in the mean flux of each
bandpasses the total uncertainties in the line strengths can be
trivially computed by propagating the errors in a manner mathematically
equivalent to \cite{jesus}. The key to our method is that the
determination of the variances in each bandpass do not rely on simple
photon statistics. In detail, the models may not provide perfect matches
within the line strength bandpasses but these detailed departures from
the models only become important when template-mismatch dominates the
errors in the velocity dispersion fitting. In the original reference for
these spectra \citep{kelsonb}, template-mismatch appears to become
significant beyond $S/N>60$ per \AA, but does not dominate the residuals
between a galaxy spectrum and the model fit to it until one has
significantly higher S/N ratios. Furthermore, by including functions in
the fit for matching the continuum, broad mismatches are improved, such
as for CN or C4668.

The accuracy of our error estimates was confirmed in two ways. First, we
performed Monte Carlo simulations, in which we degraded model spectra to
a range of $S/N$ ratios and remeasured the absorption line strengths. We
found excellent agreement between our error estimates and the scatter in
the measured line strengths about the input measurements.

The second approach we took to verify the validity of our error
estimates involved exploiting the fact that each spectrum had originally
been observed with three exposures. For each exposure we measured the
indices and their formal errors and created histograms of the
distributions of the indices about their weighted means. These are shown
in Figure \ref{fig:histo}. In the figure, Gaussians with standard
deviations of unity (the dashed lines) show the distributions one
expects if our formal error estimates are consistent with the
differences between the three separate observations per galaxy. The
solid lines show Gaussians whose standard deviations are given in each
panel and were determined empirically from the observed distribution of
line strengths from the individual exposures. Excluding Mg$b$, the mean
standard deviation of the histograms is 0.95, indicating that our
errors, on average, are over-estimates by approximately 5\%.
Unfortunately, Mg$b$ is contaminated by telluric absorption at this
redshift and our corrections to it are uncertain. Thus for that index
the variance between the measurements appears to be systematically
reduced, and, we infer, leads to an artificial reduction in $\langle
\chi^2\rangle$ for its histogram \citep[also see][for examples of the
spectra]{datapaper}. Taken together, these tests indicate that our
method for estimating the random errors in the absorption line strengths
is accurate for spectra spanning the range of signal-to-noise ratios of
these data.


\begin{figure*}[t]
\centerline{\epsscale{0.95} \plotone{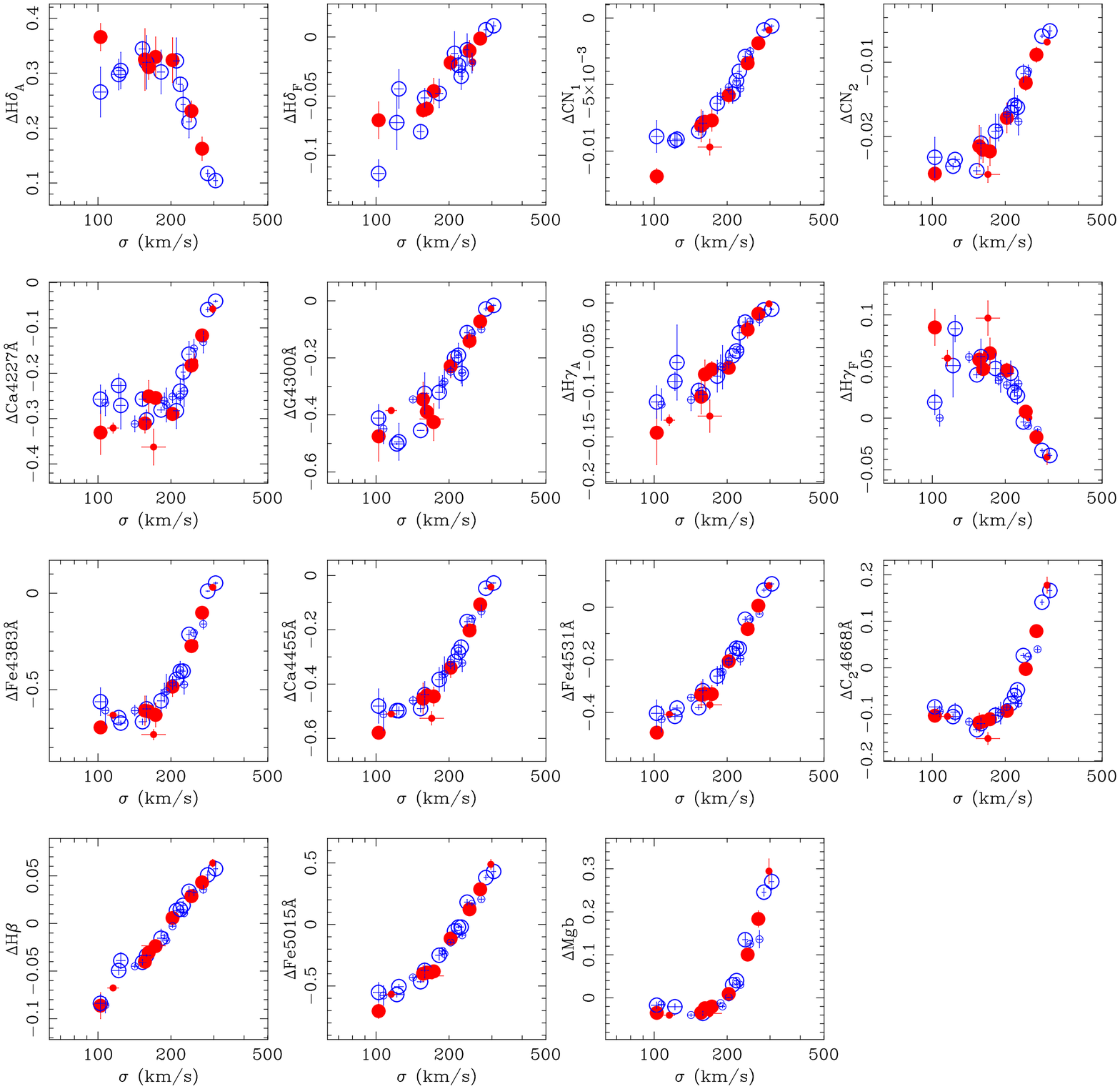}}
\caption{Broadening corrections to the blue Lick indices for the E/S0s
in CL1358+62 as a function of velocity dispersion. Symbols are as in
Fig. \ref{fig:selection}. The increased scatter at low velocity
dispersions arises because of an increase in the scatter of the ages
and/or metallicities of the \cite{bc2003} SEDs used to derive the
corrections for each of the galaxies. Much of this increased scatter is
not statistically significant, though the variations in the intrinsic
strengths of the features, i.e. variations in the underlying SEDs, are
important for indices such as H$\gamma_F$.
\label{fig:broadcor}}
\end{figure*}

\subsection{Correction for Instrumental and Doppler Broadening}
\label{sec:broad}

In spectra of galaxies, the blending of the absorption lines
significantly alters the continuum, and spectra with different intrinsic
Doppler or instrumental broadenings will yield different
pseudo-equivalent widths. Consequently, measurements of absorption line
strengths must be corrected for both the internal motions of the stars
and for the finite resolution of the spectrograph. While the system of
line strengths could have been defined by spectra with arbitrary levels
of intrinsic Doppler or instrumental broadening, the Lick/IDS system of
indices is referenced to stars (i.e., spectra with zero Doppler
broadening) observed through the IDS, for which the instrumental
resolution was characterized in \cite{wortheyhd}. It is important to
note that this characterization of the IDS's instrumental resolution has
a very strong dependence on wavelength.

The broadening corrections were computed by measuring an index, $X$,
three times for each galaxy: (1) $X|_G$, the index measured directly
from the galaxy spectrum; (2) $X|_{B(\sigma)\circ T}$, the index
measured from the template spectrum (or model SED; see previous section)
broadened to the velocity dispersion of the galaxy and at the resolution
of the galaxy's spectrum; and (3) $X|_{T_{IDS}}$, the index measured
from the template spectrum (or model SED) with a velocity dispersion of
zero and convolved to the resolution of the Lick/IDS system
\citep{wortheyhd}. Thus, each galaxy's corrected line strength
$X_{cor}|_G$ can be recovered by
\begin{equation}
X_{cor}|_G = X|_G + \biggl[ X|_{T_{IDS}} - X|_{B(\sigma) \circ T}
\biggr]
\end{equation}
There are two advantages to defining the broadening corrections in this
way. First, smoothing the galaxy spectra to low spectral resolution
dramatically, and non-trivially alters the statistics because of
non-uniform sources of noise. Second, our definition of the broadening
corrections allows for sign reversals that can occur as the
pseudo-continua change with large differences in Doppler and
instrumental broadening. In these cases, such as for the high-order
Balmer lines, multiplicative corrections \citep[e.g.,][]{jesus,trager}
can be problematic.

Because broadening corrections can be sensitive to the adopted template
SED, we adopt the best-fit \cite{bc2003} SED for $T$. By doing so the
systematic error in the broadening corrections is minimized. However,
the ages and metallicities of the best-fit \cite{bc2003} SEDs are
sensitive to the wavelength range of the fit. Therefore we derive
broadening corrections using parameters derived from fitting over $\rm
4100\AA\ < \lambda < 5000\AA$, $\rm 4200\AA\ < \lambda < 4800\AA$, $\rm
4070\AA\ < \lambda < 4800\AA$, and $\rm 4070\AA\ < \lambda < 4500\AA$,
in the restframe. We adopt the mean broadening correction from these
four fits, and we use the scatter in their broadening corrections as
estimates of the random errors in the corrections.

Fortunately, the large uncertainties in the SED ages ($\sim 0.15$ dex)
and metallicities ($\sim 0.3$ dex) are strongly (anti)correlated such
that the uncertainties in the broadening correction are small. For all
of the indices, the errors in the broadening corrections are estimated
to be at a typical level of $\pm 10\%$ {\it of the correction\/} itself;
i.e., quite small. More importantly, the uncertainties in the
corrections are typically $<3\%$ of the dynamic ranges of the line
strengths, as estimated using the 68\% widths of the distributions of
the indices.

Our broadening corrections are plotted against galaxy velocity
dispersion in Figure \ref{fig:broadcor}. At velocity dispersions
$\sigma\simlt 150$ km/s, there is an increase in the scatter in the
broadening corrections for several indices. Even though these increases
in the scatter arise because of an increase in the scatter of the
best-fit ages and/or metallicities of the \cite{bc2003} SEDs, the
increased scatter in the corrections is typically not statistically
significant, given the uncertainties in the corrections themselves. A
small number of galaxies fall off the primary loci of points for the
narrow Balmer indices H$\delta_F$ and H$\gamma_F$. These galaxies, below
the magnitude limit of the sample, do have larger values of these
indices, though the wider H$\delta_A$ and H$\gamma_A$ indices, which
will be used in our analysis below, do not show such large departures.


\subsection{Correction for Aperture Size}

In order to accurately compare samples of galaxies at different
redshifts, one must be correct for the effects of observing the galaxies
with different aperture sizes. Because this sample will be the reference
dataset in our survey of distant clusters \cite{evolpaper}, we correct
the observed line strengths to an effective aperture consistent with the
aperture used in the most distant cluster in our sample, MS1054--03, at
$z$=0.83. In \cite{datapaper} we detail our formalism for estimating the
relationship between the internal line strength gradients in distant
galaxies and the dependence of line strength measurements on aperture
size.

Using estimates of the line strength gradients for the CL1358+62 sample
galaxies, and the relationship between gradient, $\beta$, and aperture
correction, $\Delta X$, derived in \cite{datapaper}, we have corrected
the line strengths of the CL1358+62 galaxies to an aperture equivalent
to a diameter $D_{\rm ref}=10.1$ kpc at the distance of MS1054--03
($D_{ap}=1\Sec 35$, using, again, $H0=72$ km/s/Mpc, $\Omega_M=0.27$, and
$\Omega_\Lambda = 0.73$). The corrections are defined to be $\Delta X=
\beta \times f(D_{ap}/D_{\rm ref})$, where we derived $f(D_{ap}/D_{\rm
ref}) = 0.075$ in \cite{datapaper}. The corrections for aperture size
are small because the extraction aperture for the CL1358+62 galaxies was
chosen to minimize the corrections. As a result, the errors in
correcting the aperture to an equivalent aperture at the distance of
MS1054--03 are small ($\lappr 10\%$ of the correction).

The uncorrected line strengths for the full sample of galaxies in
CL1358+62, are given in several tables of \cite{datapaper}. The line
strengths of the elliptical and lenticular galaxies, corrected for both
broadening and aperture size, are reproduced here in Tables
\ref{tab:corindex-b} and \ref{tab:corindex-m}. Though these are the data
to be used in the discussions of the stellar populations below,
these corrections do not affect the conclusions drawn in this paper.


\begin{figure*}
\centerline{\plotone{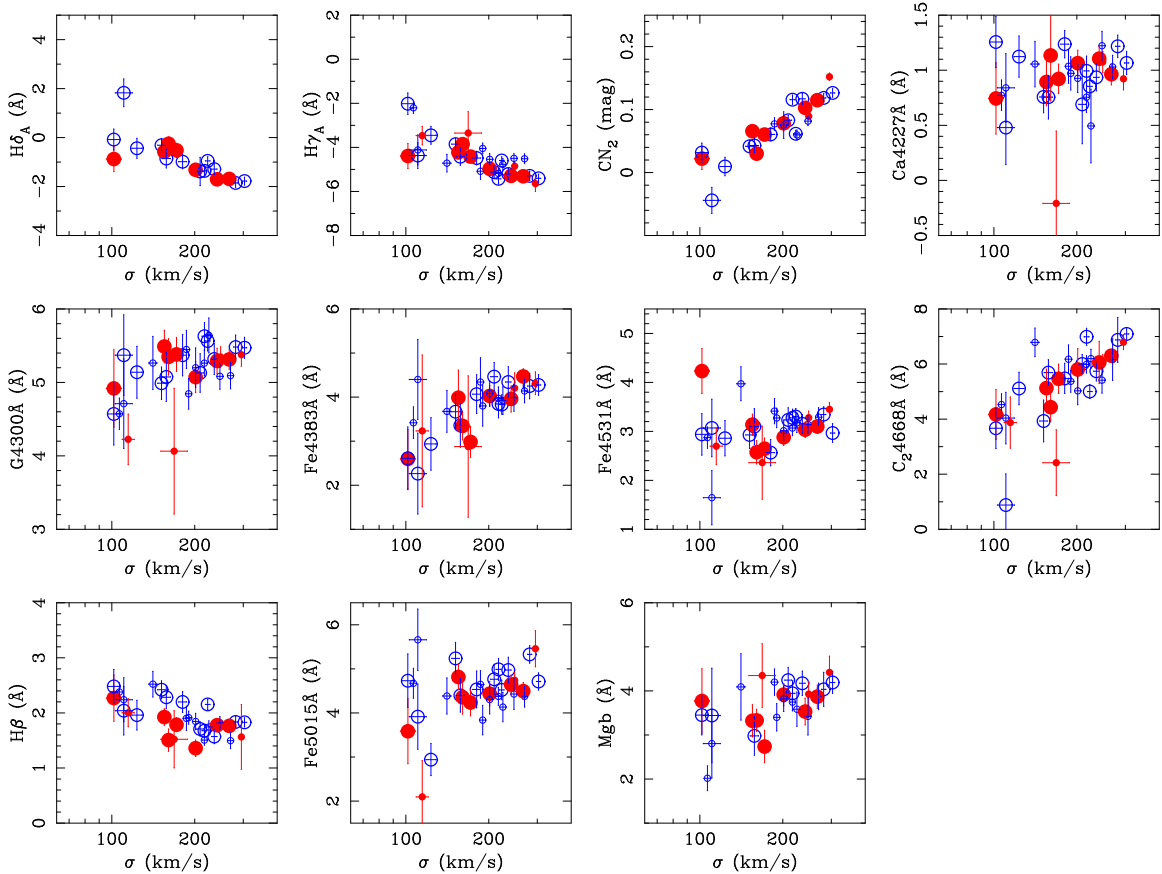}}
\caption{The line strength-line width relations for E/S0s in the
CL1358+62 sample of \cite{kelsonc}. Ellipticals are shown as filled red
circles and E/S0 and S0 galaxies as open blue circles. Note that the
measurements of Mg$b$, not used in the fitting for the stellar
population parameters, suffer from additional uncertainties arising from
poor correction to telluric absorption. The small symbols represent
those galaxies that are missing any of the eight required indices for
the fitting, and as such belong to a larger inhomogeneous sample. The
larger symbols show those galaxies which have all eight indices to be
used in the stellar population fitting.
\label{relation-early0}}
\end{figure*}

\subsection{The Line Strength-Line Width Relations}

The correlations between absorption line strengths and velocity
dispersion has become a diagnostic for directly probing galaxy evolution
\cite[e.g.][]{ziegler,kelson01,jorgensen0152}. The strengths of the
metal, molecular, and Balmer absorption features all correlate with
velocity dispersion, and the full multi-dimensional locus of early-type
galaxies in the space of velocity dispersion and line strengths has
implications for scenarios of their formation, e.g., such as the
timescale for the build-up of stellar mass. For example, the
(H$\gamma_A$+H$\delta_A$)-$\sigma$ relation was studied to $z=0.83$ by
\cite{kelson01} in an effort to directly trace the evolution of the
high-order Balmer lines, at fixed velocity dispersion, with redshift.

Before proceeding to detailed modeling of the indices, we show the
observed line strength-line width relations for the early-type galaxies
in CL1358+62 in Figure \ref{relation-early0}. Note that the E and S0
galaxies follow statistically identical line strength-line width
relations. This fact has consequences for the uniformity of their
formation and enrichment histories, and we discuss this further below
using simple models.


\section{Fitting for the Relative Stellar Population Parameters}
\label{sec:fitting}

The models published in \cite{thomas} and \cite{thomas2} contain
predictions for the Lick/IDS indices at locations in sparsely sampled
``grid'' of the six parameters: $\log t$, [Z/H], [$\alpha$/Fe] (mean
$\alpha$ enhancement), [$\alpha$/N] (nitrogen enhancement or depletion
on top of any enhancement of N as an $\alpha$ element), [$\alpha$/C]
(same as [$\alpha$/N] but for carbon), and [$\alpha$/Ca] (for calcium).
For clarity, when nitrogen is enhanced with respect to the ensemble of
$\alpha$ elements, [$\alpha$/N] decreases. As a reminder, most published
results on $\alpha$-enhancement stem from the measurement of indices
sensitive to Mg. However, oxygen is the most abundant $\alpha$ element
and will likely dominate the $\alpha$ sensitivity in the blue owing to
the equilibrium between carbon, nitrogen, and oxygen.

The dependence of these spectral indices on the controlling parameters
is non-linear and the inversion of observables to find the
luminosity-weighted mean properties of stellar populations is
non-trivial. In this section we detail the steps by which we determine
the stellar population model parameters that best fit the observed line
strengths.

Each of our galaxies in CL1358+62 has between eight and ten indices
measured from its spectrum. For every galaxy, we wish to use all of the
available data to constrain its star-formation and nucleosynthetic
history, as described by the ``simple'' models. In principle our fit of
a simple stellar population (SSP) to a given galaxy has up to six
unknowns, and of order ten observables, suggesting that the process of
inverting the observables should be straightforward. However, each Lick
index measures the combined strength of many blended features in the
spectrum of the underlying, unbroadened stellar population. These
features are {\it all\/} sensitive to the ages, mean metallicities, and
detailed abundance ratios of the stars
\citep{burstein84,wortheys,worthey,trager2000,cardiel2003,thomas,thomas2}.
To help illustrate these degeneracies, Table \ref{tab:derivs} lists
the first partial derivatives of the model predictions (at a specific
reference point, discussed below). These derivatives indicate that the
process of fitting models to observed indices will lead to significant
covariances between the fitted parameters. The derivatives given in the
table are only valid at a single location in the six dimensional space
of SSP parameters and the full dependencies of the model predictions on
the SSP parameters is nonlinear. Below we describe how the nonlinearity
of the models has been treated.

At this time we note that some derivatives are not listed. These are
explicitly assumed to be zero, and we elaborate on the implications of
this below, in the context of our method for fitting differential
stellar population parameters. The unknown derivatives have occurred
because either \cite{thomas} specifically indicated that that particular
sensitivity was negligible, or the models do not yet include the
sensitivities of these indices to those abundance ratios
\citep{tripicco95,thomas2,korn}. The next section discusses our
treatment of these shortcomings.

In principle the wavelength coverage in our spectra allows us to utilize
all of the blue Lick indices from H$\delta_A$ through Fe5015 in the
fitting of stellar population parameters for most of the galaxies. Mg$b$
was observed for many galaxies but we do not use it because of the
uncertainties introduced by the strong telluric absorption. The spectra
of some galaxies also missed portions of the H$\delta$ and CN$_1$
continua or index passbands. In theory one could derive ages and
metallicities for all of the galaxies, despite the absence of a few
random indices from some galaxies. However, the dependencies of the
parameters on the observables would not be uniform across the sample and
the resulting ages and abundances would not be homogeneous.

In an effort to construct an homogeneous set of stellar population
parameters, we opt to fit only those galaxies for which there exist
measurements of H$\delta_A$, H$\gamma_A$, CN$_2$, Ca4227, G4300, Fe4383,
Fe4531, and C4668 (H$\beta$ has been excluded from the fitting because
it can be affected by emission). Out of the original sample of 55
galaxies, 33 of which were classified as E, E/S0, or S0 by
\cite{fabric1358} in the original \cite{kelsonb} sample, we are left
with an homogeneous sample of 19 early-type galaxies for the detailed
study of stellar populations. Our primary analysis will rely on this
homogeneous sample. To help illustrate whether the fuller sample has
comparable uniformity to the smaller, homogeneous sample, the remaining
galaxies are included in some of the figures. As can be seen in Figure
\ref{relation-early0}, these additional objects obey the same relations
as the smaller subset, with no significant outliers.


\subsection{Eliminating the Zero-point}
\label{sec:recalib}

Before fitting the indices to find the relative ages and abundances, one
more issue requires resolution: systematic offsets between the system of
measured indices and the model system, defined to be equivalent to the
Lick system \citep{burstein84,trager}. There are two distinct issues
here: (1) actual differences between the observations that form the
basis of the system of model indices and the indices measured from our
spectra; and (2) unknown dependencies of indices on any of the abundance
ratios. We have attempted to correct for differences between the model
and observed systems of indices using the procedures described earlier,
but significant systematic offsets between the model predictions and the
data may remain. Such systematic differences between the systems of the
models and our data severely hamper our ability to infer the absolute
ages and chemical make-up of the stellar populations in our sample.
Separate from differences in the systems of the modeled and observed
indices, intrinsic calibration uncertainties exist for the model indices
themselves. For example, the higher order Balmer lines defined by
\cite{wortheyhd}, may have calibration, or zero-point uncertainties of
up to $\about 1$\AA\ \citep{tragerbalmer}.

The other systematic effect, that of ignored, or even unknown
dependencies of the line strength predictions on any of the various
abundance ratios, is also very important. Recently \cite{thomas2} have
determined the sensitivities of the high-order Balmer lines to
[$\alpha$/Fe], and those authors argued that past discrepancies between
ages derived from the different Balmer lines can be attributed to the
previously uncalibrated dependencies of those features on [$\alpha$/Fe].
However, other elemental dependencies, such as a dependence of
H$\gamma_A$ on nitrogen or carbon, are also likely to be important
\citep{schiavon1}.

We eliminate these two sources of systematic error by zero-pointing the
models to our data. Doing so ensures that every index yields the same
age, metallicity, etc. This serves to eliminate any zero-point
uncertainties in the indices, arising from calibration errors or from
unknown zeroth-order dependencies of the line strengths on all of the
abundance parameters. The chief penalty incurred by recalibrating the
models is the loss of the zero-points for $\log t$, [Z/H],
[$\alpha$/Fe], etc.. More specifically, the model's stellar population
parameters are no longer absolute but are relative. Fortunately we are
most interested in deriving accurate relative measures of age and
abundance, and the absolute values are simply not important for this
paper or for our survey of the properties of cluster galaxies as a
function of look-back time.

The recalibration is performed by adopting a set of stellar population
parameters for a reference point in the space of observables. This
reference point is defined by the mean massive early-type galaxy in
CL1358+62, computed, for each index, using the means of the line
strengths for the most massive E/S0s in the cluster. We adopt a mean age
of 7 Gyr \citep[i.e., $z_f=2.4$;][]{kelson01} for the reference point,
based largely in the results from previously published studies of the
evolution of E/S0 galaxies. \citep[][and
others]{kelson97,vdfp83,kelsonc,vandokkum2003,holden2004,holden2005}. We
adopt abundances of [Z/H] = 0.3; $[\alpha/{\rm Fe}]=0.2$; and
$[\alpha/{\rm N}]= [\alpha/{\rm C}]= [\alpha/{\rm Ca}]=0$ because work
on nearby ellipticals has shown that the most massive of such galaxies
typically have super-solar metallicity and $\alpha$-enhancements between
$0.2 < [\alpha/{\rm Fe}] < 0.3$ \citep[e.g.,][]{trager2000}. Within the
range of values defined by previous work on nearby galaxies, our key
results are not sensitive to changes in the adoption of these specific
parameters. In Appendix \ref{app:zpt}, we show that our key results are
insensitive to uncertainties in the adopted SSP parameters of the
recalibration.

The recalibration of the models is performed in the following fashion:
(1) model line strengths are generated for the reference SSP given
above; (2) mean line strengths are computed using the 9 E/S0 galaxies
with $\sigma> 200$ km/s; and (3) arithmetic differences between these
mean line strengths and the reference model line strengths are then used
as zero-point corrections to the models when the fitting is performed.
By doing so, we ensure that the data will always fall within the bounds
of the model grid, and the location of the $\chi^2$ minimum for a given
galaxy will be easily accessible in a downhill search beginning at the
reference point. As has been pointed out by other authors, the largest
systematic error in the location of the $\chi^2$ minimum arises from
uncertainties in the zero-points of the models (vs one's observations).
By eliminating this major, non-Gaussian source of uncertainty, the fit
for the relative population parameters can now be performed by
minimizing $\chi^2$, with the expectation that now the covariance
matrices can be used to estimate formal uncertainties.

This last point is important: even if our assumed zero-point
stellar population parameters are wrong (and they probably are), the
derivatives in the 6-dimensional model grid will be correct to first
order, and thus the relative offsets $\Delta\log t$, $\Delta$[Z/H],
..., will be correct to second order. If we had assumed that the models and
our data are on the same system, then large systematic errors in the
calibration would have translated into large uncertainties in the absolute
values of the stellar population parameters. Furthermore, because
different indices have different (or unknown) dependencies
on several of the abundance ratios, these systematic uncertainties would
have led to large errors in the ages and abundances.

Essentially such systematic uncertainties in the zero-points of our
measurements result in different absolute ages or abundances for each
index. By adjusting the zero-points of the six-dimensional grid for a
given index, one eliminates the zeroth order term in the dependence of
that index on the abundance ratios. For indices whose dependencies on
abundance ratios have been modeled, we know that the derivatives vary
slowly, and that the mixed second derivatives are small. Taken together,
the errors in the relative stellar population parameters essentially
have only second-order dependencies on the abundance ratios.

If the total systematic error between the calibration of the
observations and models had been ignored, we would not be able to
correctly estimate formal uncertainties, even for relative measures of
age and abundance. Bogus correlations between the stellar population
parameters would have appeared, and underlying correlations between
parameters might have been masked. These issues are made more explicit
in Appendix \ref{app:zpt}. All of these issues were also explicitly
verified when we incorporated the upgraded models, in which the
[$\alpha$/Fe]-dependence of the H$\gamma$ and H$\delta$ indices were
included, and our results on the differential properties of the stellar
populations did not change in any statistically significant way.

The model line strengths at the reference point are given as $X_{\rm
model}$ in Table \ref{tab:zpts}. The mean (adopted) line strengths for
the reference point are given as $X_{\rm \langle CL1358\rangle}$. The
difference, or zero-point corrections, are given by $\Delta_X$. Again,
the Appendix \ref{app:zpt} discusses the impact of ignoring this
zero-point recalibration.


\subsection{$\chi^2$ Minimization}

The derivation of the stellar population parameters for a given galaxy
is performed by searching through the six dimensional space of
parameters to find the set that minimizes $\chi^2=\sum [(X_{\rm
model}+\Delta_X-X)/\sigma_{X}]^2$ \cite[also see][]{proctor2004b},
where $X\in\{$ H$\delta_A$, CN$_2$,
Ca4227\AA, G4300\AA, H$\gamma_A$, Fe4383\AA, Fe4531\AA, and, C4668\AA\},
$\sigma_X$ is the formal error in $X$, $X_{\rm model}$ is the model
prediction for index $X$, and $\Delta_X$ is the zero-point correction
for the index, as described above. In order to linearize the
minimization of $\chi^2$, smooth approximations to the model grids are
required, and to this end the discrete model grids were replaced with
multivariate B-splines. The model indices are fit by bicubic B-splines
in both $\log t$ and [Z/H], where the knot coefficients are themselves
represented as low-order polynomial functions of [$\alpha$/Fe],
[$\alpha$/N], [$\alpha$/C], and [$\alpha$/Ca]. Constructed in this way,
the original multidimensional grids are reproduced exactly, and the
interpolating function has well-behaved derivatives everywhere the
bivariate B-spline is defined: $0 \le \log t \le 1.3$ (Gyr) and $-2 \le
{\rm [Z/H]} \le 1$. The \cite{thomas} models are defined between $-2.25
\le {\rm [Z/H]} \le 0.67$. Fortunately none of the sample discussed
here have fitted [Z/H] values outside of this range. However, any
galaxies that would have been fit using very large metallicities, e.g.,
[Z/H]$>0.67$, would have properties extrapolated from the grid.

The $\chi^2$ minimization is performed using a modified
Levenberg-Marquardt algorithm \citep{minpack}. For single stellar
populations, the mapping of parameters to observables is unique.
Therefore, we begin each search at the reference point, at old ages,
solar metallicity, and solar abundance ratios. From that portion of the
grid, the best-fit stellar population parameters are accessible by any
efficient gradient search algorithm.

The result of a fit is a set of stellar population parameters defined
relative to the mean of massive E/S0 galaxies in the cluster. These
differential SSP parameters are referred to by $\Delta\log t$,
$\Delta$[Z/H], $\Delta$[$\alpha$/Fe], .... Note that we are fitting all
of the available data, while others \citep[e.g.][]{cardiel2003} have
suggested that one can reduce the number of observables to include only
those that are most sensitive to the desired stellar population
parameters. Our methodology does not require such an analysis. Because
we have eliminated the systematic uncertainty in the zero-points, and
have reliable estimates of the formal errors in the line strengths, the
Jacobian correctly accounts for the sensitivity of the parameters to all
of the measurements (normalized by their uncertainties).


\section{The Relative Ages and Chemical Abundances}
\label{sec:fits}

The properties of the stellar populations of the 19 early-type galaxies
are found by fitting to only eight of the available observables. And
while the models do allow one to fit for six parameters, it would be
foolhardy to do so, given the uncertainties in our measurements and the
degeneracies in the models. Therefore we employ a series of constrained
models. The results of the fitting will be discussed within the context
of finding the minimum number of parameters required to define the
family of early-type galaxies.


\begin{figure*}
\leftline{\epsscale{0.57}\plotone{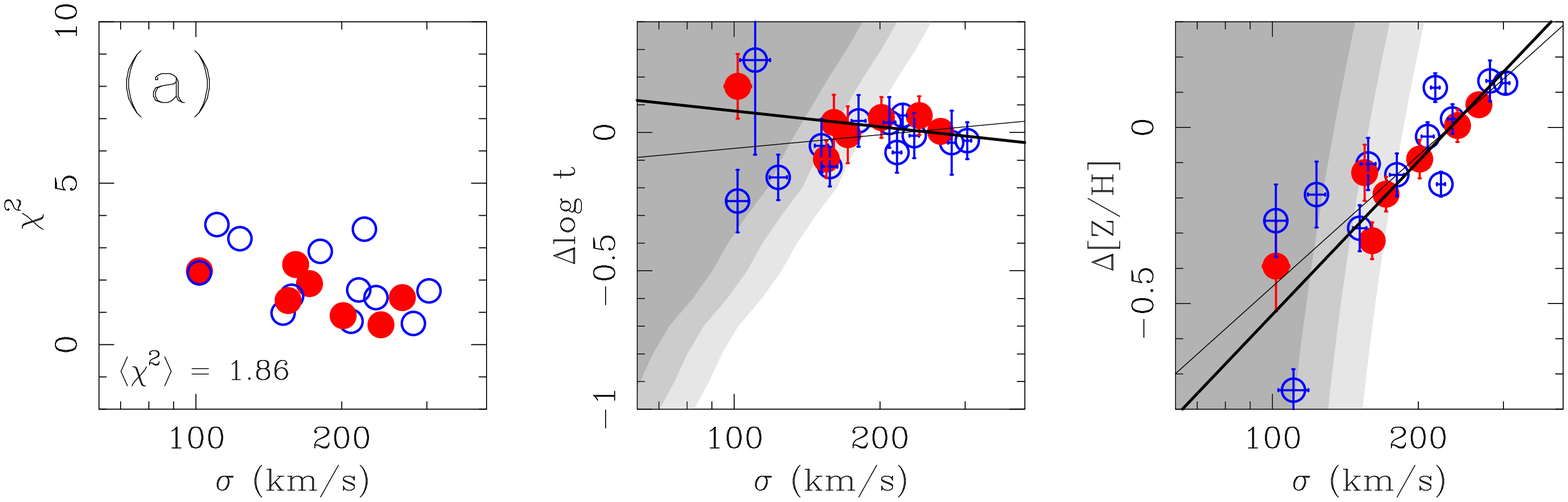}}
\leftline{\epsscale{0.76}\plotone{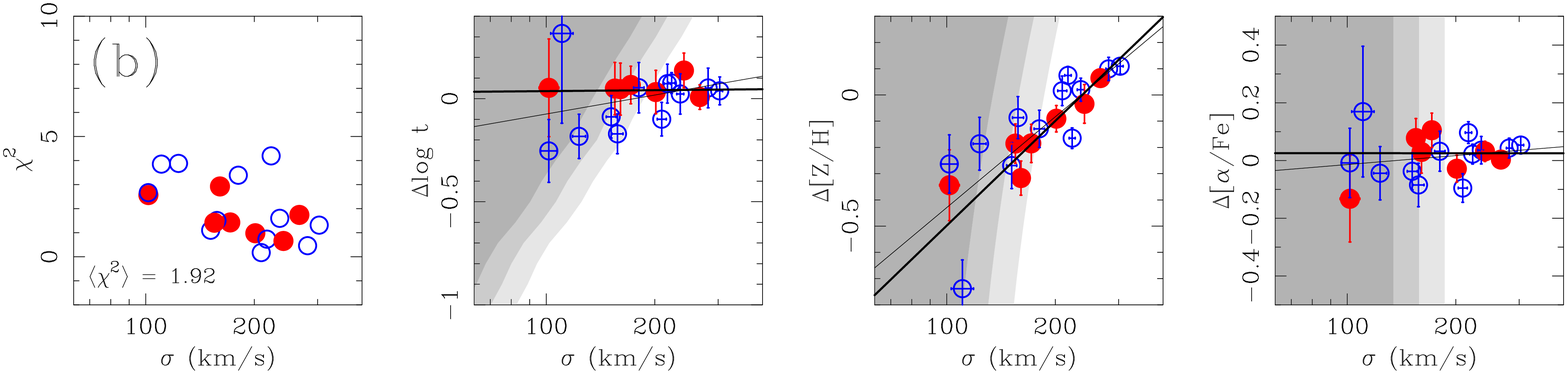}}
\leftline{\epsscale{0.95}\plotone{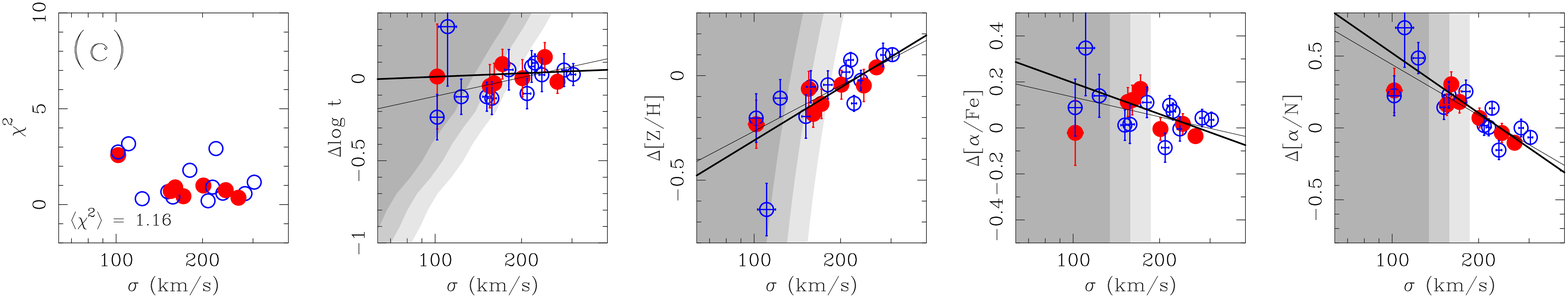}}
\caption{Stellar population parameters as a function of velocity
dispersion for ellipticals (filled red circles) and E/S0 and S0 galaxies
(open blue circles) for three different classes of models: (a) where
only relative age and metallicity are fit ($\Delta\log
t$-$\Delta$[Z/H]); (b) where relative age, metallicity ,and
$\alpha$-enhancement are fit ($\Delta\log
t$-$\Delta$[Z/H]-$\Delta$[$\alpha$/Fe]); and (c) where relative age,
metallicity, $\alpha$-enhancement, and nitrogen-enhancement are fit
($\Delta\log
t$-$\Delta$[Z/H]-$\Delta$[$\alpha$/Fe]-$\Delta$[$\alpha$/N]). Only those
galaxies with all eight observables are shown. The left-most panels show
the reduced $\chi^2$ for each galaxy, also plotted against galaxy
velocity dispersion. The mean reduced $\chi^2$ values are also given in
these panels. There are several key features to note: (1) There appears
no statistically significant correlation of age with velocity
dispersion; (2) there is a clear $\Delta$[Z/H]-$\log \sigma$
correlation; and (3) our correlations of $\Delta$[$\alpha$/Fe] with
$\log \sigma$ are marginally significant and they even appear to be
anti-correlated when nitrogen is allowed to vary; and (4) The addition
of $\Delta$[$\alpha$/N] as a fitted parameter significantly improves the
fits to the observed line strengths. The darkest gray region shows the
location of the $R=21$ mag selection cut, using the $R$-band
Faber-Jackson relation in the cluster. The lighter shades of gray show
the +1- and +2-$\sigma$ scatter about the Faber-Jackson relation. In
plotting the abundance ratios, we assume the $M/L$ ratios have zero
dependency on the specific abundance ratio in order to show the selection.
Within these gray regions, the selection effects become important and
the sample becomes non-random. As a result, we do not include those
galaxies within these biased regions when fitting for population trends
with $\sigma$. The thin and thick solid lines show the correlations
obtained when fitting using galaxies with velocity dispersions greater
than the mean dispersion at $R=21$ mag (134 km/s), and the mean plus one
standard deviation (158 km/s), respectively. The abundance trends are
insensitive to the selection biases, while apparent correlations of age
with velocity dispersion are biased strongly by those galaxies at the
limit of the sample. As a result, there is no significant evidence for a
correlation of stellar population age with $\sigma$.
\label{early-feh-age}}
\end{figure*}

\subsection{Models with Variable Ages}

Here we explore several sets of models in which the relative ages are
fit, and in which the parameters controlling the relative abundance
ratios are optionally kept frozen. These models will be used to explore
to what extent the SSP ages vary among the early-type galaxies in this
sample. Figure \ref{early-feh-age}(a) shows results when one fits only
for $\Delta\log t$ and $\Delta$[Z/H], and keeps the other four
parameters fixed. The filled red circles denote the ellipticals; the
open blue circles the E/S0 and S0 galaxies. The morphologies, again, are
taken from \cite{fabric1358}. Because the models have been recalibrated
to the data, each index should yield the same ages and abundances.
However, the different indices have different sensitivities so data
missing for particularly indices can bias the results for those
galaxies. Therefore, only those galaxies in the homogeneous sample, with
all eight observables, are shown and all discussions of correlations and
statistical variations will refer only to these galaxies.

In each of the figures for this section, we plot the best-fit relative
stellar population parameters against galaxy velocity dispersion, and
also include a panel showing the reduced $\chi^2$ of the fit for each
galaxy. In the plots of relative age, or metallicity, vs. velocity
dispersion, we show the approximate location of magnitude selection cut
using three gray tones. The darkest gray region shows the region of
velocity dispersions for galaxies below the mean dispersion at $R=21$
mag, using the Faber-Jackson relation of the full sample. The +1- and
+2-$\sigma$ scatter of the $R$-velocity dispersion correlation is
indicated by the lighter gray strips. In these regions the galaxy
selection is no longer random and the sample becomes biased. Therefore,
we can not include those galaxies in any calculation of correlations of
the parameters with redshift.

In each figure we show the effects of the selection using the thin and
thick solid lines, derived by fitting correlations using those galaxies
with $\sigma > $ 134, 158 km/s, respectively. These cuts in velocity
dispersion represent the mean dispersion, and the mean dispersion plus
one standard deviation, at $R=21$ mag. In principle, 50\% and 32\% of
the underlying sample of galaxies at the magnitude limit of $R=21$ mag
are lost. Those galaxies at the limit which are over-luminous for their
velocity dispersion, either because they are younger or metal poor, can
make it into the sample and bias the observed population trends. In
fitting for correlations, we opt not to weight using the uncertainties
because the measurement errors are significantly correlated with
$\sigma$. Uncertainties in the fitted slopes were computed using the
``bootstrap'' method (with 5000 random samples). We discuss any specific
correlations in more detail below.

For the models in which only $\Delta\log t$ and $\Delta$[Z/H] vary, the
mean $\chi^2$ per degree of freedom for the two-parameter models was
$\langle\chi^2\rangle=1.86\gg 1$, indicating that such models, in which
only these two parameters vary, are not complete descriptors of the
stellar populations. However, despite the poor quality of the fitting,
the [Z/H]-$\sigma$ relation is apparent. Discussions of the significance
of any correlations with $\sigma$ are deferred until below, when fits to
the data with lower $\langle\chi^2\rangle$ are derived.

In fitting the line strengths we find that the two parameters
$\Delta\log t$ and $\Delta \rm [Z/H]$ are insufficient for describing
the observations. Therefore, we proceed to add additional parameters,
one-by-one. Because of the previously published work incorporating
$\alpha$/Fe as a key parameter in the stellar populations of early-type
galaxies
\citep[e.g.,][]{jorgensen1999,trager2000,proctor2004a,thomas2005}, we
now allow $\Delta$[$\alpha$/Fe] to vary in the fitting. In Fig.
\ref{early-feh-age}(b), these results are shown, again plotted against
velocity dispersion. For purposes of illustration, the gray regions of
exclusion are plotted assuming that the $M/L$ ratios have zero
dependence on the abundance ratios. With $\langle \chi^2\rangle=1.92$ it
appears that $\alpha$-enhancement, by itself, does not serve to improve
the fits to the blue Lick indices. Furthermore, it suggests that
$\alpha$/Fe is not an important parameter for defining the sequence of
E/S0s in CL1358+62. The ages appear to be uniform and the
$\Delta$[Z/H]-$\log \sigma$ correlation is only slightly flatter than in
Fig. \ref{early-feh-age}(a). These models indicate that $\alpha$/Fe is a
constant in these galaxies, and not a strong function of velocity
dispersion.

These models shown in Fig. \ref{early-feh-age}(b) are directly
comparable to many previously published studies of early-type galaxies
\citep[e.g.][]{harald,trager2000,thomas2005,ziegler2005}. A linear
least-squares fit to the relative $\alpha$-enhancement as a function of
$\log\sigma$ gives a slope of $-0.01\pm0.18$ dex/dex (the thick line in
the figure), compared to the [$\alpha$/Fe] $\propto 0.33\log\sigma$
found by \cite{trager2000} or [$\alpha$/Fe] $\propto 0.28\log\sigma$
found by \cite{thomas2005}. While we do not find a statistically
significant correlation of $\Delta$[$\alpha$/Fe] in our 3-parameter
models, the difference between our result and theirs is discrepant at
the 2-$\sigma$ level (even without knowing what the uncertainties
in their slope estimates are, and with our comparatively small sample).
Further discussion comparing our constraints on [$\alpha$/Fe] with
previous results is reserved for \S \ref{sec:compare}. Nevertheless,
adding $\Delta$[$\alpha$/Fe] to the fitting of the blue indices did not
reduce $\langle \chi^2\rangle$; surely other abundance ratios are more
important.

The \cite{thomas} models allow three other abundance ratios to be
varied. In previously published analyses of line strengths, there has
been evidence for variable nitrogen abundance ratios among old stellar
populations, and \citep{burstein84,briley2002,briley2004a} among
early-type galaxies specifically \citep[e.g.][]{carretero}. We therefore
investigated the impact of the $\Delta$[$\alpha$/N] relative abundance
ratios, and the results are shown in Figure \ref{early-feh-age}(c).
First, note the significantly lower mean $\chi^2$ per degree of freedom,
$\langle \chi^2 \rangle = 1.16$, indicating that nitrogen is a key
factor in fitting the blue Lick indices (in particular the cyanogen
index). Second, there is a strong correlation of the nitrogen abundance
ratio with galaxy velocity dispersion. The origins of this correlation
will be explored in a later section.

In the models shown in Figure \ref{early-feh-age}, the slope of the
correlation of relative age with velocity dispersion is very sensitive
to selection biases. Excluding galaxies below $\sigma<158$ km/s, one
obtains $\Delta\log t\propto (0.06\pm 0.19)\log\sigma$. If one cuts the
sample at the mean dispersion at $R=21$ mag ($\sigma<134$ km/s), one
would expect that galaxies which are ``young'' for their mass/dispersion
would make it into the sample, while older galaxies drop out, thus
producing a bias of younger ages at lower velocity dispersion. In fact,
this bias appears to be present, as one finds a slope of $\Delta\log
t\propto (0.38\pm 0.18)\log\sigma$. This bias is not minimized by
selecting only those galaxies with velocity dispersions greater than the
mean dispersion at the magnitude limit of the sample, but by selecting
only those galaxies with dispersions greater than the mean plus {\it at
least\/} one standard deviation of the relation that transforms the
magnitude limit to a cut along the abscissa. The larger the scatter in
the transformation of the magnitude cut to a cut in the dependent
variable, the greater the bias in any correlation is.

Because of the above arguments, we conclude there is no significant
evidence for a correlation of stellar population age with $\sigma$, down
to the limit of our sample. Furthermore, {\it we find no significant
difference between the ages of the S0 and elliptical galaxies\/} down to
our magnitude limit. Note that this result is consistent with what was
found in our analysis of the fundamental plane \citep{kelsonc}.

The best fits to the data in which ages are allowed to vary are shown in
Fig. \ref{early-feh-age}(c). These data indicate that the stellar
populations of the early-type galaxies not only have remarkably uniform
ages at fixed velocity dispersion, but that they also follow a tight
$\Delta$[Z/H]-$\log\sigma$ correlation. The resulting distributions of
$\Delta\log t$, $\Delta$[Z/H], $\Delta$[$\alpha$/Fe], and
$\Delta$[$\alpha$/N] with velocity dispersion can be summarized by the
following:

\begin{figure*}
\leftline{\epsscale{0.76} \plotone{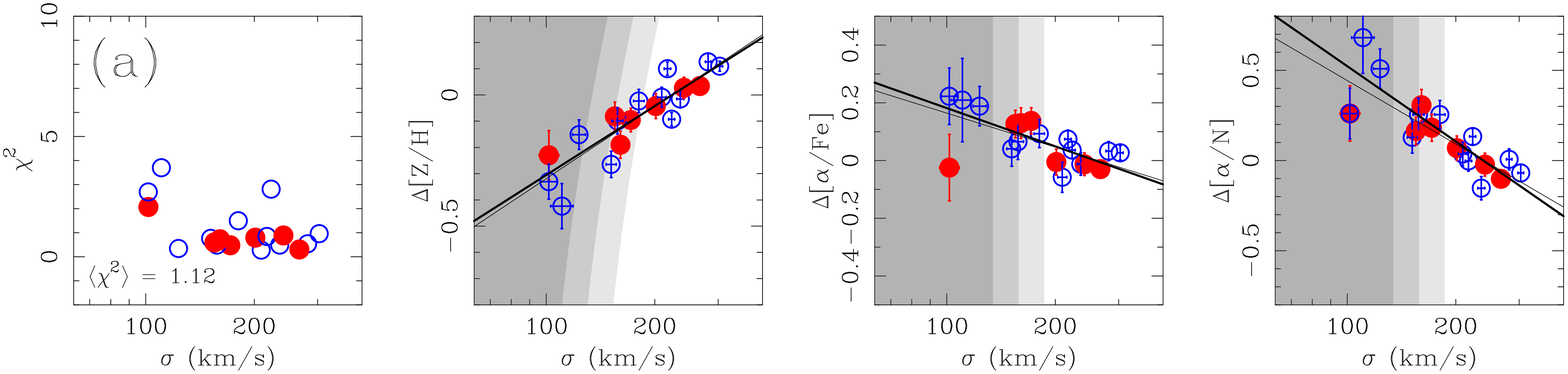}}
\leftline{\epsscale{0.95} \plotone{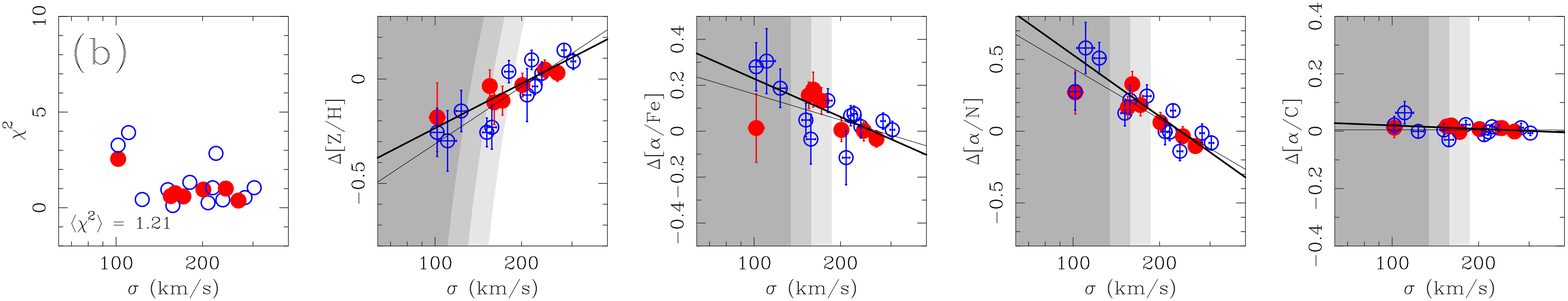}}
\leftline{\epsscale{0.95} \plotone{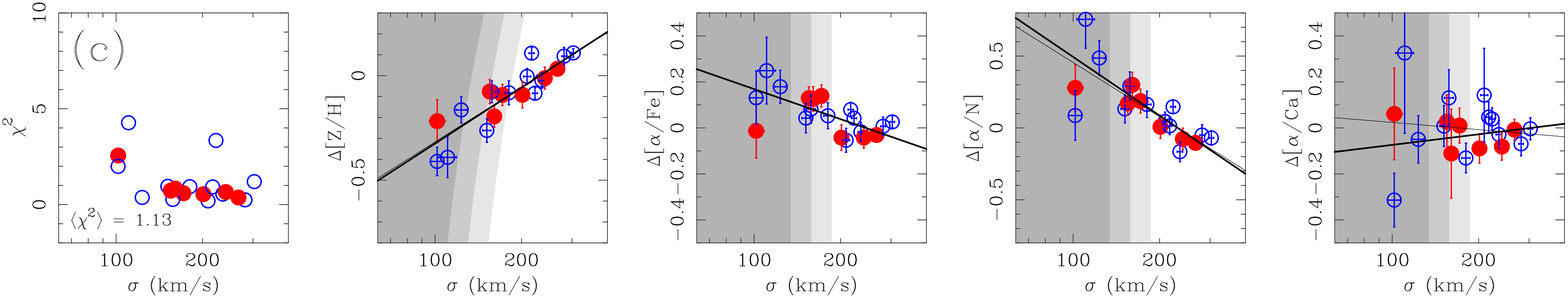}}
\caption{Same as in Figure \ref{early-feh-age} but where ages are
assumed to be constant.
(a) parameters from fitting only for
$\Delta$[Z/H], $\Delta$[$\alpha$/Fe], and $\Delta$[$\alpha$/N];
(b) parameters from fitting only for
$\Delta$[Z/H], $\Delta$[$\alpha$/Fe], $\Delta$[$\alpha$/N], and
$\Delta$[$\alpha$/C]; and
(c) parameters from fitting only for
$\Delta$[Z/H], $\Delta$[$\alpha$/Fe], $\Delta$[$\alpha$/N], and
$\Delta$[$\alpha$/Ca].
No improvements are found by allowing the Ca and C abundance
ratios to vary. We find no significant correlation of these abundance
ratios with velocity dispersion. The ramifications of the correlations
of the nitrogen and carbon abundance ratios with velocity dispersion
are discussed in \S \ref{sec:nitro}.
\label{early-feh-nitro}}
\end{figure*}

\noindent$\bullet$ The scatter in relative ages, above the limit of the
sample, is 0.06 dex. The scatter expected from the formal errors is
$0.09$ dex, suggesting that the formal errors on the relative ages have
been overestimated by a third. The observed scatter is consistent with
the low scatter in mass-to-light ratios, at fixed $\sigma$, found in
\cite{kelsonc}. There is no significant correlation of relative age with
velocity dispersion, down to the limit of the sample, consistent with
the lack of evolution of the fundamental plane slope to this redshift
\citep{kelsonc}.

\noindent$\bullet$ No difference in the relative ages and abundance
patterns of Es and S0s is seen in CL1358+62, contrary to what was seen
in Fornax by \cite{harald}. This is presumably because the Es
and S0s in our sample span similar ranges of velocity dispersions, while
the S0s in \cite{harald} all have $\sigma < 100$ km/s, and, as such,
belong to a larger, generally inhomogeneous set of early-type galaxies
\cite[e.g.][]{jfk95,cr97,kelsonc,caldwell2003}.

\noindent$\bullet$ A steep, tight correlation of $\Delta$[Z/H] with
velocity dispersion is found, with a slope of $0.86\pm 0.17$ dex/dex.
This is about twice as steep as that seen in other samples of early-type
galaxies \citep[e.g.][]{trager2000} and we investigate the implications
for the color-magnitude relation below. The scatter in $\Delta$[Z/H],
at fixed $\sigma$, is $0.06$ dex, where the scatter expected from
the formal errors is $0.05$ dex.

\noindent$\bullet$ About 30\% larger scatter is seen in
$\Delta$[$\alpha$/Fe] than the $0.04$ dex expected from the formal
errors. There appears to be a mild anti-correlation of
$\Delta$[$\alpha$/Fe] with velocity dispersion, with a slope of $-0.44
\pm 0.25$ dex/dex. Using those galaxies with $\sigma>134$ km/s, one
obtains $-0.28 \pm 0.15$ dex/dex, indicating potential sensitivity to
the selection. The inclusion of the nitrogen abundance ratio in the
fitting does have an impact on the best-fit values for
$\Delta$[$\alpha$/Fe], with many of the blue indices providing leverage
on [$\alpha$/Fe] \citep[see Table \ref{tab:derivs}
and][]{thomas,thomas2}. Without variable nitrogen abundance ratios, one
obtains $\Delta$[$\alpha$/Fe] $\propto (-0.01\pm 0.18)\log\sigma$. Such
$\alpha$-enhancement patterns differ from previously published
correlations of $\Delta$[$\alpha$/Fe] with velocity dispersion
\cite[see, e.g.][]{jesus,trager2000,mehlert2003,thomas2005}, though
earlier work relied on redder Lick indices (e.g. Mg$b$), employed models
in which the nitrogen was not free to vary. We discuss the implications
of these $\Delta$[$\alpha$/Fe] estimates on the Mg-$\sigma$ relation
below.

\noindent$\bullet$ A steep anti-correlation of $\Delta$[$\alpha$/N] with
velocity dispersion is found, with a slope of $-1.36 \pm 0.32$ dex/dex.
Using those galaxies with $\sigma>134$ km/s, one obtains $-1.17 \pm
0.22$ dex/dex. Given the results for $\alpha$/Fe, we can restate the
correlation of nitrogen enhancement with velocity dispersion as $[{\rm
N/Fe}] \propto (0.94\pm 0.24)\log \sigma$, or $[{\rm N/H}] \propto
(1.78\pm 0.25)\log \sigma$. The implications of the apparent
nitrogen-$\sigma$ correlation will be discussed further below, though at
a basic level the nitrogen abundance ratio appears to play an important
role in dictating the behavior of many blue Lick indices, especially
given its impact on the fits for [$\alpha$/Fe]. The scatter in
$\Delta$[$\alpha$/N], at fixed $\sigma$, is 0.08 dex where the scatter
expected from the errors is 0.06 dex.


\subsection{Models with Variable Carbon and Calcium}

The results of the fits in the previous section indicate that the ages
of the early-types show no statistical variation down to the limit of
our selection. Therefore we now fit models in which we forced
$\Delta\log t=0$. Fixing the relative ages does not lead to an increase
in reduced $\langle \chi^2\rangle$, consistent with the scatter in
relative ages arising from the formal errors alone. Note that freezing
the ages introduces no changes to the correlations of $\Delta$[Z/H],
$\Delta$[$\alpha$/Fe], or $\Delta$[$\alpha$/N] with $\log \sigma$
discussed in the previous section. This is shown graphically in Figure
\ref{early-feh-nitro}(a). Note, too, that the mean $\chi^2$ per degree
of freedom has actually decreased, providing further proof that fitting
for the ages does not contribute any statistically significant
information about the stellar populations.

Since the relative ages can be constrained without affecting the other
parameters, we now explore whether $\Delta$[$\alpha$/C] and
$\Delta$[$\alpha$/Ca] have any significant variation among the E/S0s in
CL1358+62. Figures \ref{early-feh-nitro}(b) and (c) show that these
additional variables do not improve the quality of the fits to the blue
Lick indices, and that the carbon and calcium abundance ratios show no
statistically significant variation or correlation with $\log \sigma$.
We also verified that the inclusion of these abundance ratios has no
effect on the correlations of $\Delta$[Z/H], $\Delta$[$\alpha$/Fe], and
$\Delta$[$\alpha$/N] with velocity dispersion shown above.

Given the extreme sensitivity of the C4668 index to both [Z/H] and
[$\alpha$/C], we find these results quite remarkable. First, the lack of
a correlation of the carbon abundance ratio with velocity dispersion
indicates that the broadening corrections were accurate (fitting the
uncorrected line strengths leads to an anti-correlation of
$\Delta[\alpha/{\rm C}]$ with $\log \sigma$). In the context of the
steep correlation of nitrogen abundance with velocity dispersion,
constant carbon abundance ratios have important ramifications, to be
discussed in the next section. Lastly, though the constraints on the
calcium abundances are not strong, we see no evidence for a trend of
$\Delta$[$\alpha$/Ca] (or, by extension, $\Delta$[Ca/Fe]) with velocity
dispersion, contrary to what has been published by previous authors
\citep{sagliaCa,cenarro,falc,mich2003,thomas-ca}.


\begin{figure*}
\centerline{\plotone{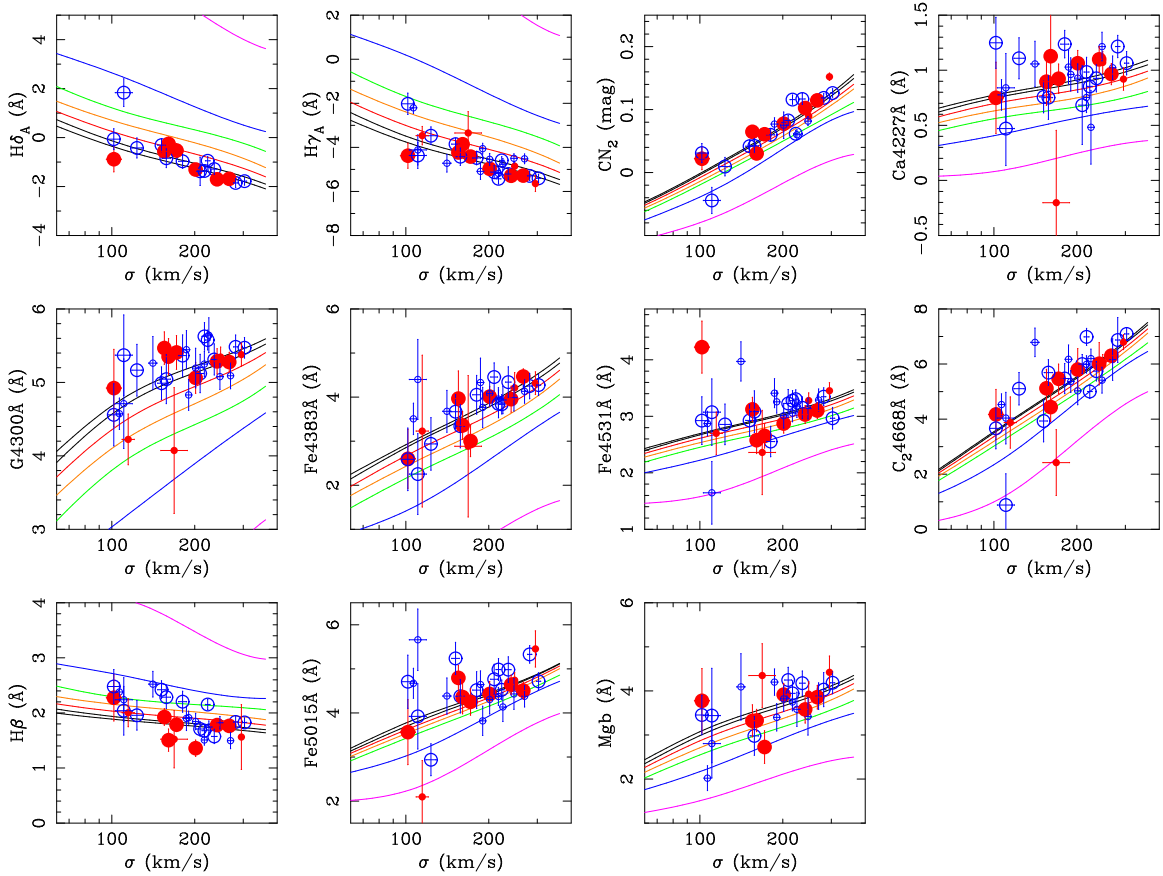}}
\caption{The line strength-line width relations for E/S0s in CL1358+62.
Symbols are as in Figure \ref{relation-early0}. The solid lines show
curves of constant age and constant $\alpha$/Fe, defined by projecting
the $\Delta$[Z/H]-, and $\Delta$[$\alpha$/N]-$\sigma$ relations. The
ages range from 1 Gyr to 7 Gyr (violet, blue, green, orange, etc.) in
increments of 1 Gyr. Several items are worth noting: (1) while H$\beta$,
Fe5015, and Mg$b$ were not used in the fitting, the fitted correlations
of stellar population parameters with velocity dispersion reproduce the
observed H$\beta$-, Fe5015-, and Mg$b$-$\sigma$ relations in CL1358+62
(though Mg$b$ has some residual contamination of telluric absorption);
(2) we also show all the E/S0s in the sample, including those galaxies
that did not have observations of all of the fitted indices and were not
included in the analysis of the homogeneous set of 19 E/S0s discussed in
the text; and (3) the slopes of the line strength-line width relations
are sensitive to the ages of the galaxies, even when there is no
variation of age with velocity dispersion. The \cite{thomas}
\citep[and][]{thomas2} stellar population models do an excellent job of
describing the relative properties of the E/S0s in this cluster.
\label{relation-early1}}
\end{figure*}

\section{Summary of the Model Fitting and Implications}
\label{sec:summary}

The models explored in this paper indicate: (1) that the ages of the
elliptical and lenticular galaxies in CL1358+62 are remarkably
homogeneous, consistent with the analysis of the fundamental plane of
these galaxies \citep{kelsonc}; and (2) the stellar populations in these
galaxies obey tight, well-defined [Z/H]-, and [$\alpha$/N]-$\sigma$
relations. Together these trends, when propagated through stellar
population models, should reproduce all of the observed early-type
galaxy scaling relations in CL1358+62 and to this end we show the
observed line strength-line width relations for the sample in Figure
\ref{relation-early1}. Note that the figure also includes those galaxies
that did not satisfy our criterion of having measurements for all of the
indices fit in the previous discussions, showing a total of 33 E/S0s.
This full sample is inhomogeneous in the sense that any galaxy's stellar
population parameters would not necessarily derive from an identical set
of indices. Note that along with the eight fitted indices we have also
included H$\beta$, Fe5015\AA, and Mg$b$ in the figure.


\subsection{The Line Strength-Line Width Projections}
\label{sec:compare}

In the individual line strength-line width diagrams we project the
metallicity- and nitrogen-$\sigma$  relations, assuming no correlation
of $\alpha$/Fe with velocity dispersion. The lines show loci of constant
ages ranging from 1 Gyr to 7 Gyr (the latter essentially the defined age
for the reference point used in recalibrating the models to the data;
see \S \ref{sec:recalib}). The oldest isochrones clearly reproduce the
observed line strength-line width relations well in CL1358+62, and it is
worth noting that the slopes of the isochrones are predicted to evolve
with age. This ``evolution'' arises from the nonlinearity of the model
grids. The implication is that any observed evolution in the slopes of
line strength-line width relations with redshift should not necessarily
be taken as proof that stellar population ages vary along the sequence
of early-type galaxies.

\begin{figure*}
\centerline{\plotone{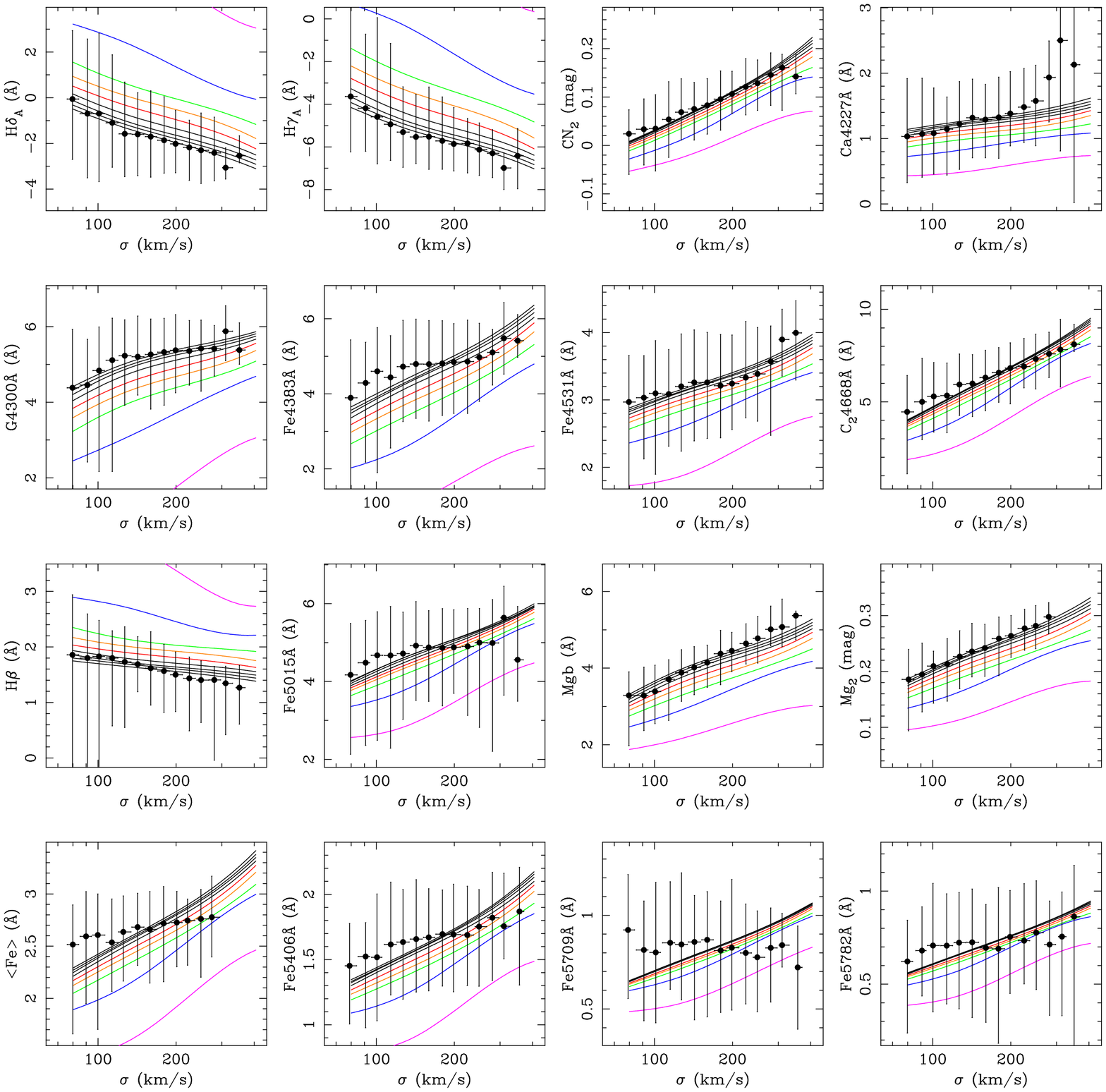}}
\caption{A comparison of isochrones following the [Z/H]- and
[$\alpha$/N]-$\sigma$ relations of the text with selected data from
the survey of nearby clusters by \cite{nelan2005}. The points show the
median line strengths of those galaxies with OIII equivalent widths
less than the median of the sample, binned in velocity dispersion in
units of 0.05 dex. The vertical bars show the bounds within which 90\%
of the sample are distributed. The predicted loci of galaxies agree well
with the distribution of galaxies in the nearby clusters for many of the
indices. The data of \cite{nelan2005} show mildly flatter line
strength-line width relations for the narrow atomic indices Fe4383,
Fe5015, $\langle\rm Fe\rangle$, Fe5406, Fe5709, and Fe5782 in contrast
to the excellent match to Fe4383 in Fig. \ref{relation-early1}.
These discrepancies, as well as the divergence of Ca4227 at large
dispersion, may be due to differences in the treatment of the broadening
corrections as these are the narrowest indices.
\label{fig:local}}
\end{figure*}

Previous analyses of redder Lick indices explicitly relied on Mg line
strengths, and strong correlations of [$\alpha$/Fe] with velocity
dispersion have been inferred from Mg$_2$ or Mg$b$
\citep[e.g.][]{terlevich,jesus,gorgas,trager,kuntschner2001,sanchez2004,thomas2005}.
However, our data simply do not reproduce the long-standing correlation
of $\alpha$-enhancement with $\sigma$. The blue line strength-line width
relations of the E/S0s in CL13562 are fit well with no correlation, or
perhaps even with a mild anti-correlation [$\alpha$/Fe] with $\log
\sigma$. Furthermore, we note that despite explicitly excluding
H$\beta$, Fe5015, and Mg$b$ from the fitting, their line strength-line
width relations are well described by the projections of the
abundance-velocity dispersion relations found with the bluer indices
(though our Mg$b$ observations have larger scatter due to uncertainties
in the corrections for telluric absorption).

\begin{figure*}
\centerline{\epsscale{0.76}\plotone{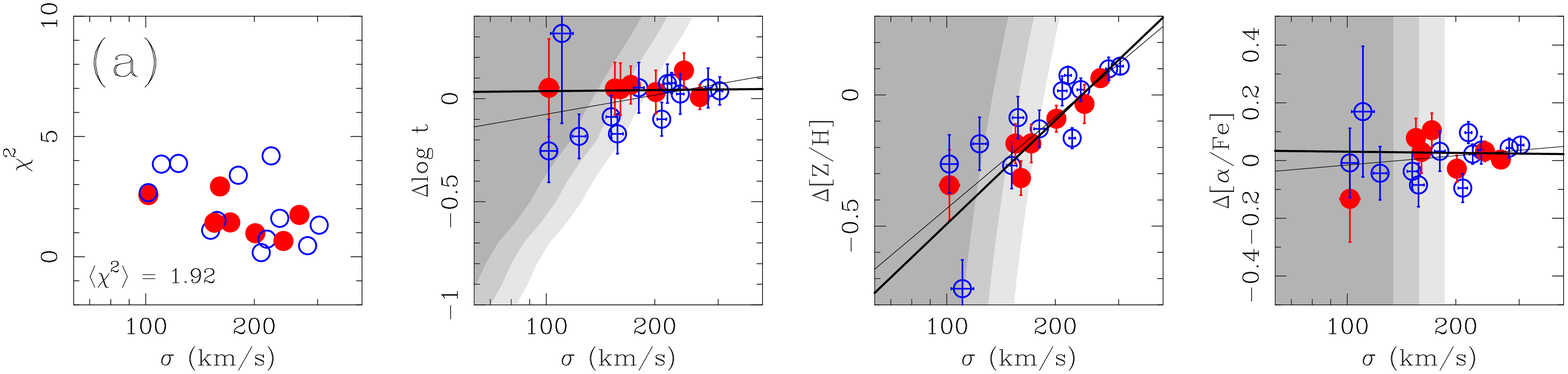}}
\centerline{\epsscale{0.76}\plotone{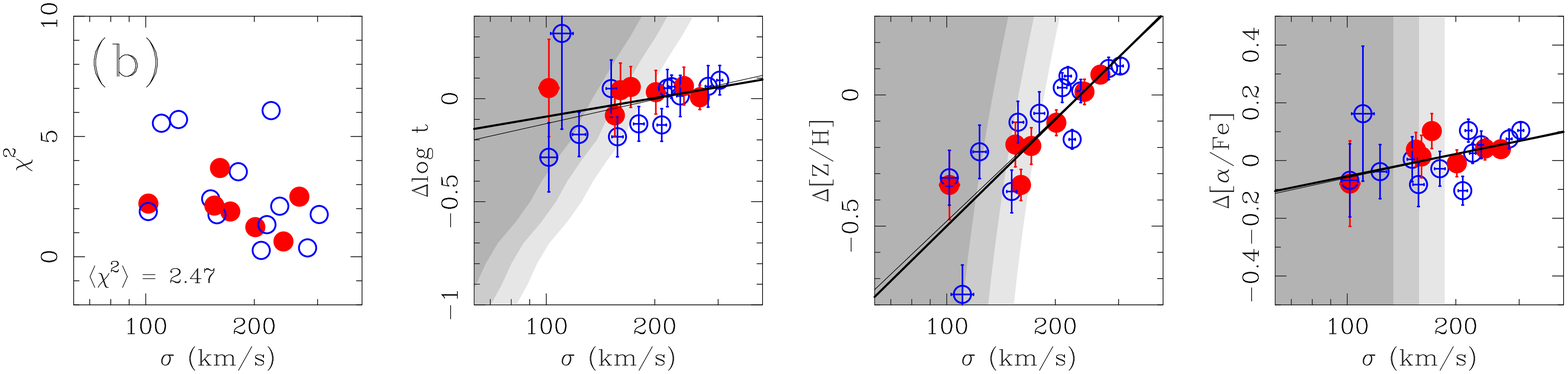}}
\caption{A comparison of the best-fit parameters $\Delta\log t$,
$\Delta$[Z/H], $\Delta$[$\alpha$/Fe] (with no additional nitrogen
enhancement) using (a) the line strengths corrected for instrumental and
Doppler broadenings using the method in \S \ref{sec:broad};
(b) the line strengths determined using a traditional approach to
correcting the effects of instrumental and Doppler broadenings. Subtle
differences in the treatment of the broadening corrections lead to
different inferred correlations of $\alpha$-enhancement with velocity
dispersion.
In (a), the best-fit slope is
$[\alpha/{\rm Fe}]\propto (-0.01\pm 0.18)\log \sigma$, while
in (b), the best-fit slope is
$[\alpha/{\rm Fe}]\propto (+0.27\pm 0.13)\log \sigma$, more consistent
with previously published results.
\label{early-feh-age-alpha}}
\end{figure*}

Our estimates of $\alpha$/Fe have implications for the origin of the
Mg$_2$-, Mg$b$-, and $\langle \rm Fe\rangle$-$\sigma$ relations
\cite[e.g.]{thomas2005}. Using the models we can recast the metallicity-
and nitrogen-$\sigma$ correlations as Mg-$\sigma$ relations and test
whether our results are consistent with previous measurements of its
slope (though the nitrogen has no impact on the Mg indices in the
current incarnations of the models). After passively evolving these
relations to the present epoch, we find Mg$_2\propto (0.23\pm 0.05)\log
\sigma$, in excellent agreement with that found nearby \citep[typical
slopes nearby range from 0.17 to 0.23 mag/dex;][]{worthey2003}. For
Mg$b$, the \cite{thomas} and \cite{thomas2} models predict that Mg$b$ is
about twice as sensitive to changes in $\alpha$/Fe, so a comparison to
Mg$b$ should provide a stronger test of the consistency of our results
with previously published data. We infer Mg$b\propto (2.8\pm 0.7)\log
\sigma$, consistent with the Mg$b\propto 2.8\log\sigma$ typically seen
\cite[see][]{worthey2003}. Fitting the Mg$b$ data of the full
inhomogeneous sample directly, we obtain Mg$b\propto (2.1\pm 0.8)\log
\sigma$ (see Fig. \ref{relation-early1}).

Because our results are consistent with the Mg$b$-$\sigma$ (and
Mg$_2$-$\sigma$) relation, we now explore potential sources of the
discrepancy over $\alpha$/Fe. In Figure \ref{fig:local} our [Z/H]- and
[$\alpha$/N]-$\sigma$ correlations have been propagated through the
models and plotted over data taken in a survey of nearby clusters from
\cite{nelan2005}. Those authors observed more than 4000 red-sequence
galaxies in 93 low-redshift clusters. Because of the large number of
galaxies in their survey, and the low typical signal-to-noise ratio of
their spectra, we bin their data in intervals of 0.05 dex, and use only
those galaxies with minimal detected OIII emission. The vertical bars
mark the bounds within which 90\% of the galaxies fall. The points show
the median line strength within each bin.

Overall the comparison between the \cite{nelan2005} data and the
predicted line strength-line width relations is quite good. However
there are particularly interesting discrepancies: the relations for
Fe4383, Fe5015, and $\langle\rm Fe\rangle$, Fe5406, Fe5709, Fe5782. For
these indices, our best-fit stellar population trends produce steeper
slopes than observed in the \cite{nelan2005} data. Because $\langle\rm
Fe\rangle$ has historically been one of the primary constraints on
metallicity, we conclude that the shallow slope of the $\langle\rm
Fe\rangle$-$\sigma$ relation of \cite{nelan2005} is characteristic of
the kinds of relations that have led to flatter [Fe/H]-$\sigma$
relations than indicated by our data. Such a flat [Fe/H]-$\sigma$
relation invariably necessitates a positive correlation of [$\alpha$/Fe]
with velocity dispersion in order to match the slope of the
Mg$b$-$\sigma$ relation. As can be seen in Fig. \ref{relation-early1},
our Fe4383-$\sigma$ relation (and Fe5015-$\sigma$, though with larger
scatter) has a slope that agrees with the projection of the models,
thus, in tandem with the other indices, negating the need for a
significant correlation of $\alpha$/Fe with velocity dispersion. Clearly
the primary source of the discrepancy in $\alpha$/Fe comes from the
narrow Fe indices. Note, too, the divergence of Ca4227 at large values
of $\sigma$.

These discrepant indices are some of the narrowest in the Lick system,
and because the remaining indices (those that require very small
corrections) are matched very well by our abundance patterns, one must
suspect that the discrepancy in $\alpha$/Fe-$\sigma$ arises from
differences in the treatment of the corrections for instrumental and
Doppler broadening. These narrow indices have had large, systematic
corrections for Doppler broadening applied to them by \cite{nelan2005}
\cite[see, e.g.,][for plots of such corrections as functions of velocity
dispersion]{poggianti2001}. Systematic errors in such corrections can
modify the inferred slopes of line strength-line width relations.

\begin{figure*}
\centerline{\epsscale{0.65} \plotone{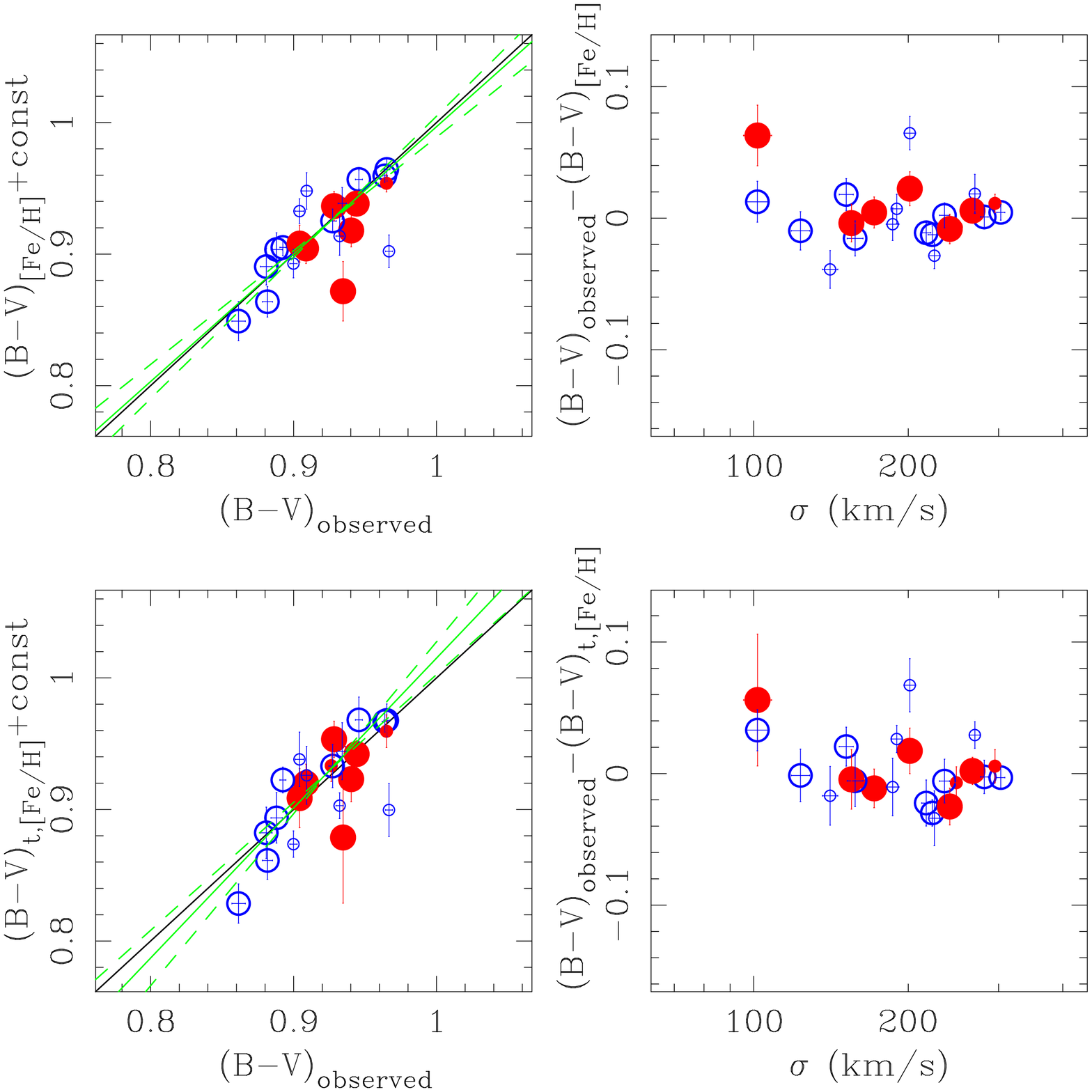}}
\caption{A comparison of observed restframe $(B-V)_{\rm observed}$ colors
\citep{kelsona,kelsonc} with those predicted using the best-fit stellar
population parameters \citep{bc2003}. The top panels show predicted
colors derived when age was kept constant. The bottom panels show the
predicted colors when age was allowed to vary. In the left-hand panels
the black solid line indicates a slope of unity. When only the
abundances were allowed to vary, the best-fit slope is
$(B-V)_{\hbox{\small[Z/H]}} \propto (0.97\pm 0.11) (B-V)_{\rm
observed}$, shown by the green solid line (where the 1-$\sigma$
uncertainties are shown using the green dashed lines). When age is
allowed to vary one obtains $(B-V)_{\hbox{\small t,[Z/H]}} \propto
(1.14\pm 0.16) (B-V)_{\rm observed}$, also shown in green. The effects
of including age are small because of (1) the statistically
insignificant variations in age in the sample, and (2) the large
covariances between [Z/H] and $\log t$ and the age-metallicity
degeneracy in $(B-V)$ colors. The differences between the observed and
predicted colors are shown plotted against velocity dispersion in the
right-hand panels. The residuals in the colors are remarkably small,
consistent with the formal uncertainties.
\label{fig:color}}
\end{figure*}

We tested whether the different $\alpha$/Fe-$\sigma$ correlation was an
inherent property of the sample or whether the treatment of the
corrections played a role. For the first test we derived line strengths
using the traditional method, i.e. first smoothing the spectra to the
resolution of the IDS, then measuring the indices, and finally applying
multiplicative corrections to these measurements. These velocity
dispersion corrections were constructed using an old, metal-rich
\cite{bc2003} model SED. These line strengths were propagated through
the machinery to derive new stellar population parameters and the
results are compared in Figure \ref{early-feh-age-alpha} to the
parameters found earlier. As can be seen in Fig.
\ref{early-feh-age-alpha}(b), we recovered $\Delta$[$\alpha$/Fe]
$\propto (+0.27\pm 0.13)\log \sigma$, in much better agreement with
previously published correlations. In this test, the corrections were
computed using a single template. Using the best-fit \cite{bc2003} SED
for each galaxy instead of a single template, we find no significant
change in the slope of the inferred correlation (though the scatter is
larger).

Given this discrepancy, we performed a second test of our broadening
corrections by smoothing the galaxy spectra to have intrinsic
broadenings equivalent to the Lick/IDS resolution with zero velocity
dispersion, i.e. by subtracting, in quadrature, the velocity dispersion
from the IDS resolution. We remeasured the indices in these spectra and
compared them to our broadening-corrected indices derived earlier. We
found no statistically significant systematic differences between these
measurements and the measurements corrected in \S \ref{sec:broad}, for
the line strengths of those galaxies with velocity dispersions less than
the IDS resolution at the wavelength of a given central passband.
Therefore we conclude that our treatment of the broadening corrections
has not introduced any gross systematic errors as a function of velocity
dispersion.

Lastly, it is worth re-emphasizing that our diagnostics of
$\alpha$-enhancements differ from those traditionally used. For example,
our spectral coverage does not provide direct constraints on magnesium
abundances. The blue indices should have stronger (though indirect)
sensitivities to oxygen abundances, primarily through the CNO
equilibrium in the atmospheres of cool giants. Because oxygen is the
dominant $\alpha$ element, our absence of a correlation of $\alpha$/Fe
with velocity dispersion may be a more accurate descriptor of the
behavior of $\alpha$ elements than indicated by Mg alone (though we note
the excellent match to the slopes of published Mg-$\sigma$ relations).
In other words, our $\alpha$/Fe results may simply reflect more on a
uniformity in O/Fe than in Mg/Fe, not to mention our empirically
constrained uniformity in the carbon and calcium abundance ratios.

While the broadening corrections appear to play an important role in the
determinations of $\alpha$/Fe in the 3-parameter fitting, these
corrections become less important when including the nitrogen
enhancement. When using the line strengths corrected for broadening via
the traditional method, we obtain similar results to those shown in
Figure \ref{early-feh-age}(c) for the 4-parameter fitting of age,
metallicity, $\alpha$/Fe, and N/$\alpha$. In \S \ref{sec:nitro} the
inferred nitrogen-$\sigma$ correlation will be discussed in more detail,
coming to the conclusion that it has a natural origin.

In the next section we show that the observed colors of the galaxies
match those predicted by the stellar population parameters that we have
derived.

\begin{figure*}
\centerline{\epsscale{0.35} \plotone{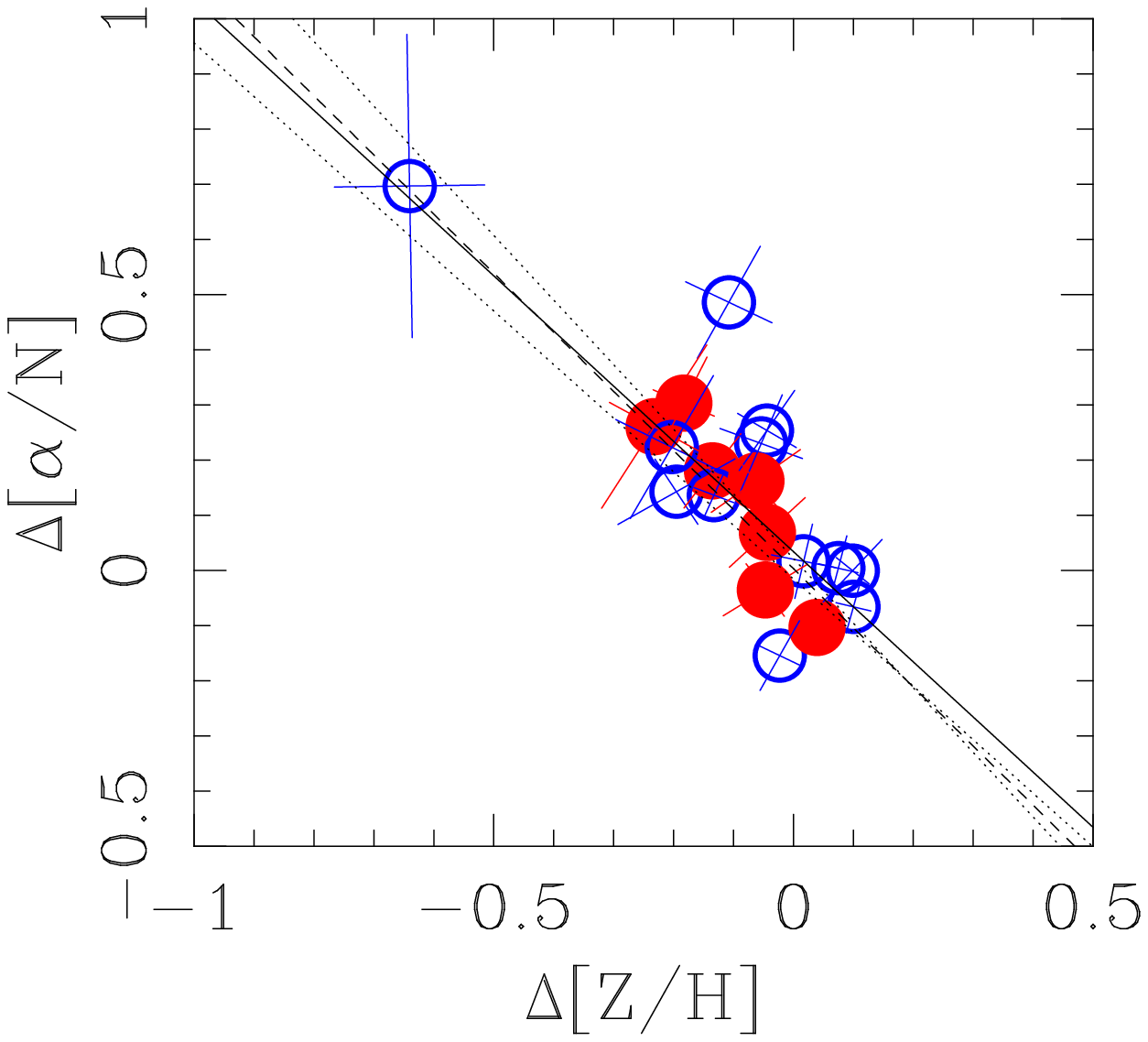}}
\caption{The correlation of $\Delta [\alpha/{\rm N}]$ and $\Delta [{\rm
Z/H}]$. For simple models of chemical enrichment
\citep{koppen1999,henryCN,koppen2005} in which abundances are dominated
by secondary nitrogen, one expects the relation to follow $\Delta
[\alpha/{\rm N}] = - \Delta [{\rm Z/H}]$, shown by the solid line.
We fit for the best-fit linear relation, using those galaxies with
$\sigma>134$ km/s, by minimizing the squares of the residuals parallel
to both axes of the error ellipses simultaneously. The resulting
best-fit slope is $(-1.07\pm 0.10)$ dex/dex, shown by the dashed and
dotted lines. The agreement between the stellar population parameters
and the prediction of simple chemical evolutionary models is striking.
Note that this slope was derived using galaxies above the limit of the
selection, and thus excluded, for example, galaxy 493 with the lowest
metallicity and highest nitrogen abundance.
\label{fig:nitro-feh}}
\end{figure*}


\subsection{The Color-Magnitude Relation}

The [Z/H]-$\sigma$ correlation has strong implications for the
color-magnitude relation. We explore these using the WFPC2 data
published by \cite{vdcm33} and \cite{kelsona}. These images were
convolved with Gaussians to mimic the seeing in our spectroscopy. Then
$\rm (F606W-F814W)$ colors were measured within circular apertures
similar to the apertures used for the spectral extractions. The colors
were transformed to restframe Johnson colors, $(B-V)_{\rm observed}$. In
this section we compare these observed colors to those predicted by the
\cite{bc2003} stellar population models using ages and metallicities
from our analysis. In order to eliminate any systematic errors in the
photometric transformation, in the calibration, or in the uncertainties
in the absolute ages and metallicities, we applied an offset to the
colors computed using the biweight estimator of \cite{beers}, much like
what was done earlier for the line strength fitting.

Figure \ref{fig:color} shows a comparison of these predicted colors with
the observed $(B-V)_{\rm observed}$ colors. When the ages were kept
constant, we refer to the predicted colors as $(B-V)_{\hbox{\small
[Z/H]}}$. When the ages were allowed to vary in the fit, the predicted
colors are referred to as $(B-V)_{\hbox{\small t,[Z/H]}}$.

In the left-hand panels one finds that both sets of predicted colors
match the observed colors well. Simple fits to the comparisons yield
$(B-V)_{\hbox{\small[Z/H]}} \propto (0.97\pm 0.11) (B-V)_{\rm
observed}$, and $(B-V)_{\hbox{\small t,[Z/H]}} \propto (1.14\pm 0.16)
(B-V)_{\rm observed}$, and these are shown using the green lines. The
solid black line indicates a slope of unity. Because colors suffer from
an age-metallicity degeneracy similar to that which plagues the
absorption line strengths, the comparison of the colors only serves as a
sanity check --- stellar populations are indeed redder as they become
older or more metal-rich, and in the correct proportion to the changes
in spectroscopic ages or metallicities. In other words, one gains no
{\it additional\/} leverage on the stellar population parameters by
fitting $(B-V)$ colors with the absorption line strengths.

Simply put, the fitted ages and metallicities reproduce the
color-velocity dispersion relations remarkably well. In the right-hand
panels we plot the color residuals against $\sigma$, and the very low
scatter, $\pm 0.02$ mag, is consistent with the formal uncertainties.

Had the ages and metallicities from Figure \ref{early-feh-age-alpha}(b)
been used, the slopes of the correlations between the predicted and
observed colors would have resulted in slopes of
$(B-V)_{\hbox{\small[Z/H]}} \propto (1.25\pm 0.10) (B-V)_{\rm
observed}$, and $(B-V)_{\hbox{\small t,[Z/H]}} \propto (1.50\pm 0.17)
(B-V)_{\rm observed}$, with significantly larger scatter. This is
because a significant, positive correlation of $\alpha$-enhancement with
velocity dispersion decreases the slope of the [Z/H]-$\sigma$
correlation to the point that the color-velocity dispersion relation can
no longer be matched.

In summary, the colors predicted by our population parameters are in
excellent agreement with the observed colors, when measured in the
appropriate aperture. This remarkable consistency between the colors and
the spectroscopic properties of these galaxies also implies that dust is
not an important contributor to the colors of early-type galaxies.


\subsection{The Nitrogen Relation}
\label{sec:nitro}

In fitting the \cite{thomas} models to the blue line strengths, we found
a strong apparent correlation of nitrogen enhancement with velocity
dispersion. However, there appears to be no associated correlation (or
anti-correlation) with the carbon abundance ratio. The lack of
concordance between the nitrogen and carbon enhancements was not
expected. Taken together with our inability to reproduce the previously
published correlation between [$\alpha$/Fe] and $\log\sigma$, our
results would seem to indicate either fundamental problems with the
stellar population models, or something truly fundamental about the
formation and enrichment histories of galaxies.

Certainly, uncertainties in the models are important \cite[see e.g.][for
a review of unexplained C and N variations among globular
clusters]{gratton2004}. The nitrogen- and carbon-sensitive indices, are
affected by mixing from first dredge-up, by other mixing processes as
well \cite[again, see][for a thorough summary and detailed
references]{gratton2004}, and by the unknown abundances of oxygen, which
are a wholly separate canard on their own
\citep{bensby2004,proctor2004b,thomas2005}. Recent work on understanding
the carbon and nitrogen abundances in Galactic and globular clusters
continue to confound
\citep[e.g.][]{aoki1997,bellman2001,briley2002,briley2004a}, though it
appears likely that both pollution by early generations of
intermediate-mass asymptotic giant branch (AGB) stars
\cite[e.g.]{briley2004b} and additional mixing processes
\cite[e.g.][]{boothroyd1999} occur. Taken all together, one cannot rule
out if the deduced N/C correlation with velocity dispersion is an
artifact arising from an absence of processes that modify the
atmospheric abundance ratios in the models, processes that are
themselves sensitive to metallicity
\citep[e.g.][]{charbonnel1994,bensby2004}.

Recalibration of the models to the data eliminated several systematic
errors, as illustrated in Appendix \ref{app:zpt}, and as a result, the
{\it relative\/} measures of age and abundance are probably quite
reasonable. However, the N/C ratios may yet be reflecting limitations in
the stellar libraries and fitting functions
\citep{worthey,thomas,thomas2}, in that the parameter space of second
derivatives of the indices has not yet been fully explored and modeled
\citep{tripicco95,korn}. Sufficiently large, but unmodeled,
non-linearities in the underlying sensitivities of the indices could be
invoked to eliminate the nitrogen abundance trends but the efforts to
calibrate the models using the integrated light of globular clusters
\citep[e.g.][]{schiavon1,thomas,proctor2004b} provide some confidence in
our results.

It is worth noting that the more common line strength-line width
relations, such as Mg-$\sigma$, appear to be well-described without any
need for a correlation of $\alpha$-enhancement with velocity dispersion,
all while the blue Lick indices are fit quite well. This
self-consistency did, however, come at a cost: that of throwing away the
zero-points of the models and the absolute measures of abundance and
age. Had we not done so, the {\it reduced\/} $\chi^2$ on individual
galaxies increase by factors of 2--7! Taken together with the large
systematic uncertainties shown in the appendix, we have much greater
confidence in the relative measures of abundance derived above than in
estimates of absolute abundances.

The steep correlation of the nitrogen abundance ratio with velocity
dispersion has interesting implications, but perhaps not for the most
obvious reason. Naively one might invoke a physical correlation between
gravitational potential and the timescales and/or efficiency of
star-formation \citep[e.g.][]{henry,wortheyreview,henryCN} given the
sensitivities of nitrogen abundances to these parameters for closed-box
models. However, the underlying correlation for nitrogen is probably not
with velocity dispersion {\it per se\/} but with mean metallicity:
secondary nitrogen is produced in intermediate-mass AGB stars
\cite[e.g.][]{henryCN} and the yields are metallicity-dependent
\citep{marigo1998}. Closed-box (and partially closed-box) models of
chemical enrichment with modest or high metallicities have N/O ratios
dominated by secondary nitrogen such that $\Delta {\rm [N/O]} \propto
\Delta {\rm [O/H]}$ \cite[e.g.][]{koppen1999,henryCN,koppen2005}. The
dependency of the yields on metallicity sets the absolute zero-point of
the logarithmic relation, but one expects the relation between relative
nitrogen abundance ratios and mean metallicities to have unity slope.
From the tables in \cite{henryCN}, the carbon yields of AGB stars do not
appear to be particularly sensitive to metallicity; presumably this is
why our carbon abundance ratios do not correlate with velocity dispersion
(or metallicity) in the same way that nitrogen does.

The hypothesis that the $\alpha$/N-$\sigma$ relation is due to secondary
nitrogen can be tested because [N/O] is predicted to correlate with
[O/H]. Within the context of our data, [$\alpha$/N] is simply a
surrogate for [O/N]. Furthermore, even though oxygen abundances cannot
be directly measured, our findings of $\Delta [\alpha/{\rm Fe}]\sim 0$
(or $\Delta [{\rm O/Fe}]\sim 0$) imply that we can approximate $\Delta
{\rm [Z/H]} \sim \Delta {\rm [O/H]}$. In Figure \ref{fig:nitro-feh}, the
relative nitrogen abundance ratios $\Delta [\alpha/{\rm N}]$ against
$\Delta [{\rm Z/H}]$. Finding the best-fit relation between $\Delta
[\alpha/{\rm N}]$ and $\Delta [{\rm Z/H}]$ is complicated by the
correlated errors in the derived abundances so we chose to minimize the
squares of the residuals parallel to both axes of the error ellipses
simultaneously. We find that the slope of the correlation is $(-1.07\pm
0.10)$ dex/dex, shown by the dashed and dotted lines in the figure.
Using other approaches to fit the correlation led to similar results. We
therefore conclude that our data are consistent with the hypothesis that
the [$\alpha$/N]-$\sigma$ correlation arises from the dependence of
yields of secondary nitrogen on metallicity.

The stellar populations of the E/S0 galaxies in CL1358+62 have
essentially been reduced to a one-parameter family, obeying a
$[{\rm Z/H}]$-$\sigma$ relation. While the presence of (secondary)
nitrogen in proportion to the metallicity indicates that the galaxies
experienced star-formation over an extended period of time, absolute
calibration of the abundances are required in order to derive
constraints on the timescales of astration. Even without absolute
calibrations, it bears repeating: the populations of both lenticulars
and ellipticals have statistically identical relative ages, span similar
ranges of metallicity, obey identical [Z/H]-$\sigma$ relations, and
show strong evidence for the presence of secondary nitrogen.

Our results conclusively show that cluster S0s did not form their stars
at significantly later epochs than cluster ellipticals of the same mass,
and the presence of secondary nitrogen indicates that both Es and S0s
formed from ancestors with extended star-formation histories. If our
$\alpha$/Fe results are correct, then this star-formation must have been
sufficiently extended, within a given galaxy, such that the relative
proportions of by-products of Type Ia to Type II supernovae are
effectively constant, and not strongly dependent on galaxy mass.
Furthermore, if the gravitational potential is an important factor in
the retention or ejection of enriched materials \citep[e.g.][and much
subsequent work]{burke1968}, then material from {\it all\/} energetic
sources of enrichment have roughly comparable efficiencies of ejection
from the host galaxies, either because of generally low probabilities of
ejection, or because the enriched interstellar gas is well-mixed before
it either forms the next generations of stars or escapes from its host.

This produces an apparent contradiction with the observed evolution in
the morphology-density relation \citep{morph2,postman2005}, in which
fewer S0s exist at higher redshift \citep[though see][]{holden2006},
even though the timescales and mechanisms with which S0s form must be
very similar to the timescales mechanisms for the formation of
ellipticals of the same mass.

Ultimately, the chemical signatures, and any evolution thereof, must be
placed in the context of cosmological models. In $\Lambda$CDM, galaxies
form hierarchically, through multiple generations of mergers
\citep{kauffman1993,kauffman1999,romeo2005,lanzoni2005,tran2005}.
Unfortunately, multiple major merger events can have a large impact on
the observed scaling relations \citep[e.g.][]{kaviraj}, such as the line
strength-line width relations analyzed here. Furthermore,
\citep[e.g.][]{juneau} directly measured the star-formation rates of
galaxies with these masses over redshifts of $0.8 < z < 2$ and found
that their star-formation has essentially ceased by $z\simgt 1.5$. In a
subsequent paper we will try to address these issues by analyzing the
stellar populations of both early- and late-type galaxies in a sequence
of five clusters from $z=0$ to $z=0.83$, referencing them to the
galaxies presented in this paper. Together it is hoped that these five
clusters can provide a more detailed picture of the formation histories
of cluster galaxies than can be derived from a snapshot at a single
epoch.

\section{Conclusions}
\label{sec:conclusions}

We have measured absorption line strengths for galaxies in the cluster
CL1358+62 using the spectra published in \cite{kelsonb}. A homogeneous
population of early-type galaxies has been analyzed, with the full
selection of galaxies in \cite{kelsonb} to be discussed in a subsequent
paper. The largest source of systematic errors that plague the analysis
of absorption lines has been eliminated by matching the stellar
populations models \citep{thomas,thomas2} to the mean observed line
strengths of the most massive early-type galaxies in CL1358+62.
Furthermore, not only does a recalibration of the models allow for
accurate differential measurements of the stellar population parameters,
but the topology of the $\chi^2$ minimum can be used to accurately
estimate formal errors.

Using only those E/S0 galaxies which have accurate measurements of eight
blue Lick/IDS indices (H$\delta_A$, H$\gamma_A$, CN$_2$, Ca4227, G4300,
Fe4383, Fe4531, and C4668), we fit for the {\it relative} measures of
age and chemical abundances. The fitting has resulted in homogeneous
sets of stellar population parameters relative to the reference point
used in resetting the model zero-point.

The key results can be summarized as follows:

\noindent (1) The populations of the E and S0 galaxies are statistically
identical in their age and abundance patterns, down to the magnitude
limit of $R=21$ mag, consistent with what was found using the
fundamental plane \citep{kelsonc}. Beyond the magnitude limit, selection
biases become important.

\noindent (2) The observed scatter in relative ages is 0.06 dex, where
the scatter expected from measurement errors is 0.09 dex. The scatter in
ages is consistent with that inferred from the color-magnitude relation
\citep{vdcm33}, and from the fundamental plane \citep{kelsonc}.

\noindent (3) We find a tight correlation between relative metallicity
and velocity dispersion, with a slope of $0.86\pm 0.17$ dex/dex. The
scatter about this relation is 0.06 dex, where the scatter expected from
the formal errors is 0.05 dex. The observed $B-V$ colors are also
consistent with the inferred ages and metallicities.

\noindent (4) The scatter in relative $\alpha$-enhancement is only 30\%
larger than expected from the formal errors alone. When the nitrogen
abundance ratio is allowed to vary, we find a mild anti-correlation of
$\Delta$[$\alpha$/Fe] with $\log \sigma$, but with a significance less
than $<2\sigma$. While we find little evidence for a significant
correlation (or variation) of [$\alpha$/Fe] with velocity dispersion,
the slope of the local Mg-$\sigma$ relation can be reproduced by our
$\Delta$[Z/H]-$\sigma$ correlation. \cite{thomas2005} argued that
Mg$b$-$\sigma$ arises mostly from the correlation of metallicity with
velocity dispersion. Our results go beyond that and suggest that no
variation of $\alpha$/Fe is required. The discrepancy between our
conclusions and previously published work on this topic arises from our
steeper [Z/H]-$\sigma$ relation. This discrepancy may be due to the
treatment of the broadening corrections, in particular for the narrow
Lick iron indices, though more detailed analysis is required to verify
this hypothesis. Previous work \cite[e.g.][]{trager2000} indicated that
cluster E/S0s form a three-parameter family of objects, with [Z/H] and
$\alpha$/Fe correlated with velocity dispersion. Historically, models
with variable ratios of Type I and Type II supernovae and subsequent
galactic winds have been invoked to explain such a complex family of
objects. Elimination of the correlation of $\alpha$-enhancement with
velocity dispersion would greatly simplify models of E/S0 formation.

\noindent (5) We find that the enhancement of nitrogen is strongly
correlated with velocity dispersion, while carbon abundance appears to
show no variation. This N-$\sigma$ correlation has not been observed
before. This new correlation originates either in deficiencies in the
stellar population models, or implies new physics to be incorporated
into our understanding of E/S0 formation and evolution. The most likely
explanation is one in which the nitrogen abundance ratios do not arise
from mechanisms specifically tied to $\sigma$, but that the
nitrogen-enhancement is correlated with metallicity. In other words, the
nitrogen is secondary in origin, and the progenitors of E/S0 galaxies
experienced significantly extended star-formation.

These data form the only sample of early-type galaxies in which these
particular abundance patterns have been found, so far. In all other
respects the sample shows the uniform age distribution and a steep
correlation of mean metallicity with velocity dispersion inferred from
other diagnostics.

While models which include variable abundance ratios
\citep[e.g.][]{trager2000,thomas,thomas2} are relatively immature, their
development has been encouraging, and the data are already well-fit by
such models. The blue Lick indices appear to be powerful diagnostics of
stellar population ages and provide excellent, self-consistent fits to
the metal lines, even if the resulting abundance ratios run counter to
much of the previous literature. Despite these discrepancies, the
observed line strength-line width relations of nearby galaxies,
published by \cite{nelan2005}, are recovered, with some notable
exceptions. Those relations that match well do bolster our faith in
these findings. However, nearly all of the iron line strength-line width
relations are not matched well, and the published data on nearby
galaxies have had large multiplicative Doppler corrections applied. We
suggest that perhaps past treatment of these corrections may be in
error. That the $(B-V)$ colors are consistent with our stellar
population parameters not only reinforces our conclusions, but also
indicates that dust is not an important component in the optical SEDs of
early-type galaxies.

The results on enhancements (over Fe) of $\alpha$, nitrogen, and carbon
have important implications. While it remains to be seen whether the
carbon- and nitrogen-sensitive indices are being modeled correctly, the
nitrogen enhancement-velocity dispersion correlation agrees very well
with the correlation of [N/Z] with [Z/H] expected to arise from
secondary production of nitrogen by AGB stars. Taken together, our
differential stellar population parameters suggest that these massive
early-type galaxies form a family with only a single parameter: galaxy
velocity dispersion (or possible mass).

The uniform ages and simple trends of metallicity with velocity
dispersion appear to contradict the observation that fewer S0s exist at
higher redshift \citep{morph2,postman2005}. We conclude that the
timescales and mechanisms with which the populations in S0s form are
statistically identical to that of ellipticals of the same mass.
Early-type galaxies, now shown to have significant self-enrichment and
extended star-formation histories, must be reconciled with hierarchical
formation \citep[e.g.][]{kauffman1993} in a $\Lambda$CDM universe. The
uniform ages and tight mass-metallicity relation may not be compatible
with significant merging and histories of extended star-formation.

In a subsequent paper, we will employ our machinery for measuring these
age and abundance patterns in a range of clusters from $z=0$ to
$z=0.83$. With comparisons of the properties of cluster galaxies over a
range of redshifts, we will test the universality of some of the
conclusions presented here, quantify changes in the stellar population
parameters with cluster redshift \citep[see,
e.g.,][]{kelson01,jorgensen0152,barr2005}, and attempt to reconcile the
above results with a coherent picture of the formation and evolution of
galaxies in rich clusters.

\acknowledgements

Appreciation is expressed to many people for fruitful discussions while
this project has persisted, including B.Holden, S.~M. Faber, A. Dressler, A.
Oemler, S.~C. Trager, F. Schweizer, W.~F. Freedman, R. Schiavon, B.
Poggianti, R. Davies, \& P. McCarthy. This research was
funded, in part, by grants HST-GO-09772.01-A and NASA grant NAG5-7697.
Lastly, we thank the referee, N.~Cardiel, for his thoroughness and
hist thoughtful comments, all of which served to make the text more
precise and more readable.

\appendix


\section{The Effects of the Systematic Recalibration}
\label{app:zpt}

\subsection{Differential versus Absolute Population Parameters}

In this section we show several figures to illustrate the effects of
recalibrating the models to our data. In particular we compare figures
shown earlier to variants in which this step is ignored.

\begin{figure*}
\leftline{\epsscale{0.95}\plotone{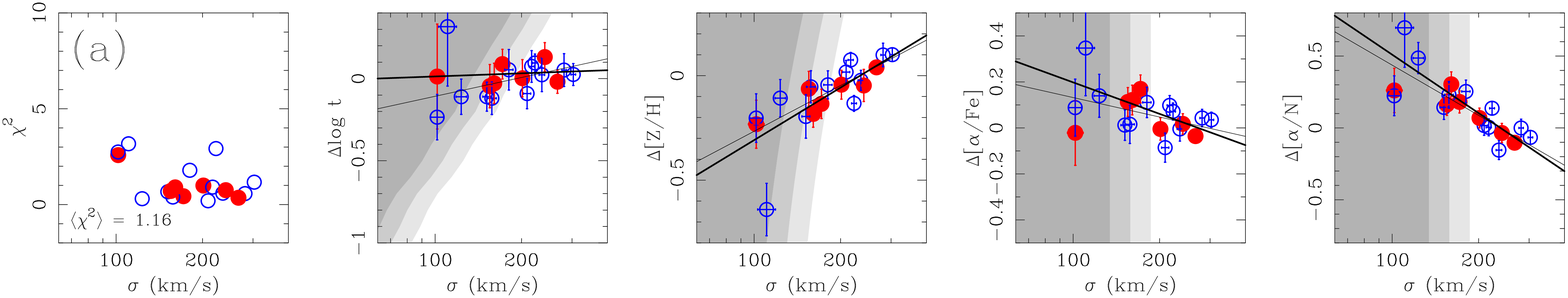}}
\leftline{\epsscale{0.95}\plotone{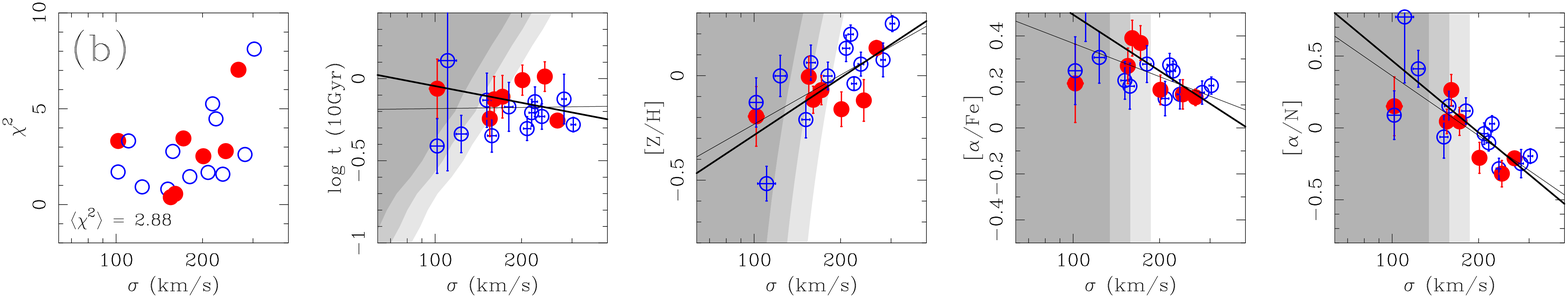}}
\caption{The effects of ignoring the model recalibration on the stellar
population parameters as functions of velocity dispersion for
ellipticals (filled red circles) and E/S0 and S0 galaxies (open blue
circles). Only those galaxies with all eight observables are shown. (a)
$\Delta\log t$-$\Delta$[Z/H]-$\Delta$[$\alpha$/Fe]-$\Delta$[$\alpha$/N]
models. (b) $\log t$-[Z/H]-[$\alpha$/Fe]-[$\alpha$/N] models. The
left-most panels show the reduced $\chi^2$ for each galaxy, also plotted
against galaxy velocity dispersion. The mean reduced $\chi^2$ values are
also given in these panels. While the general trends shown in Figure
\ref{early-feh-age} are clearly visible, the scatter about the
correlations is 30--40\% larger when the models are not recalibrated to
the data.
\label{abs-early-feh-age}}
\end{figure*}

Figure \ref{abs-early-feh-age} shows the parameters of the best-fit
models, in which age, metallicity, $\alpha$-enhancement, and nitrogen
enhancement are allowed to vary. Figure \ref{abs-early-feh-age}(a) is
identical to Fig. \ref{early-feh-age}(c). In Figure
\ref{abs-early-feh-age}(b) we show the parameters derived when the
zero-point uncertainties are ignored and the models and data are assumed
to be on the same system. Similar trends are seen in both the
differential and absolute parameters. However the absolute parameters,
derived when the recalibration is skipped, have 30--40\% larger scatter.
These systematic uncertainties appear not to destroy the general trends
of the SSP parameters with velocity dispersion but artificially inflate
the inhomogeneities in the inferred properties of the stellar
populations.

\begin{figure*}
\centerline{
\hbox{\epsscale{0.25} \plotone{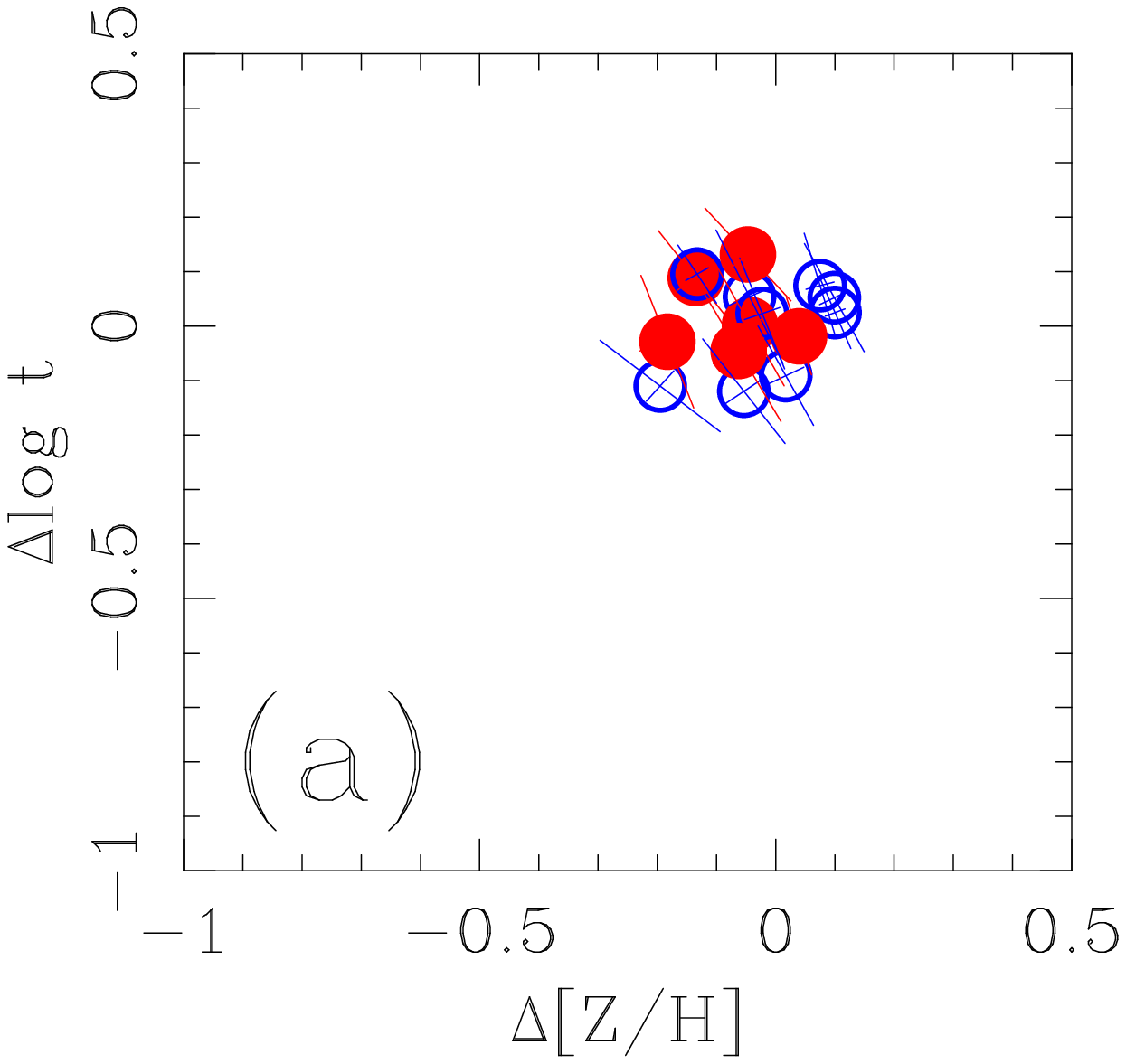}}
\hbox{\epsscale{0.25} \plotone{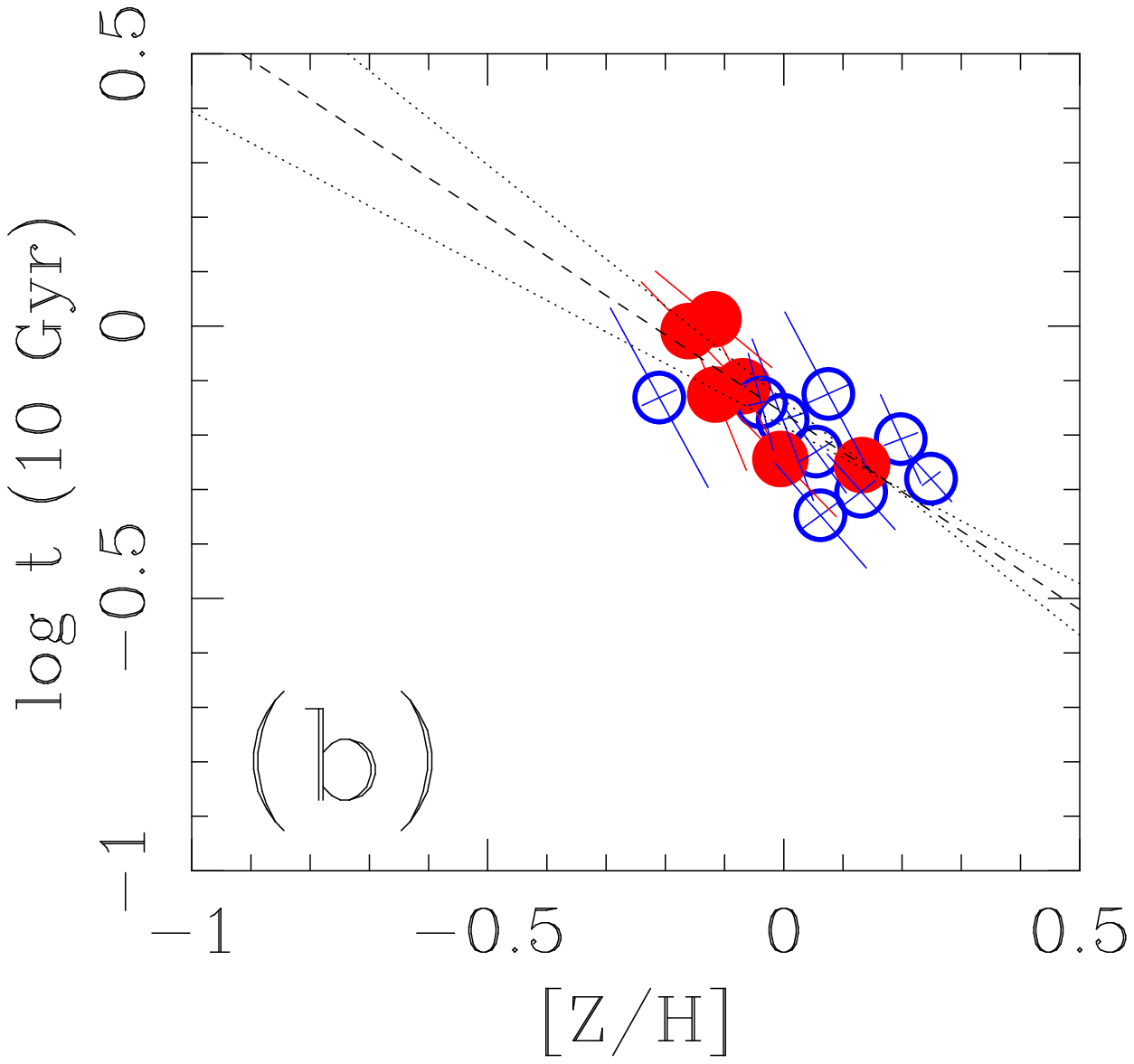}}}
\centerline{
\hbox{\epsscale{0.25} \plotone{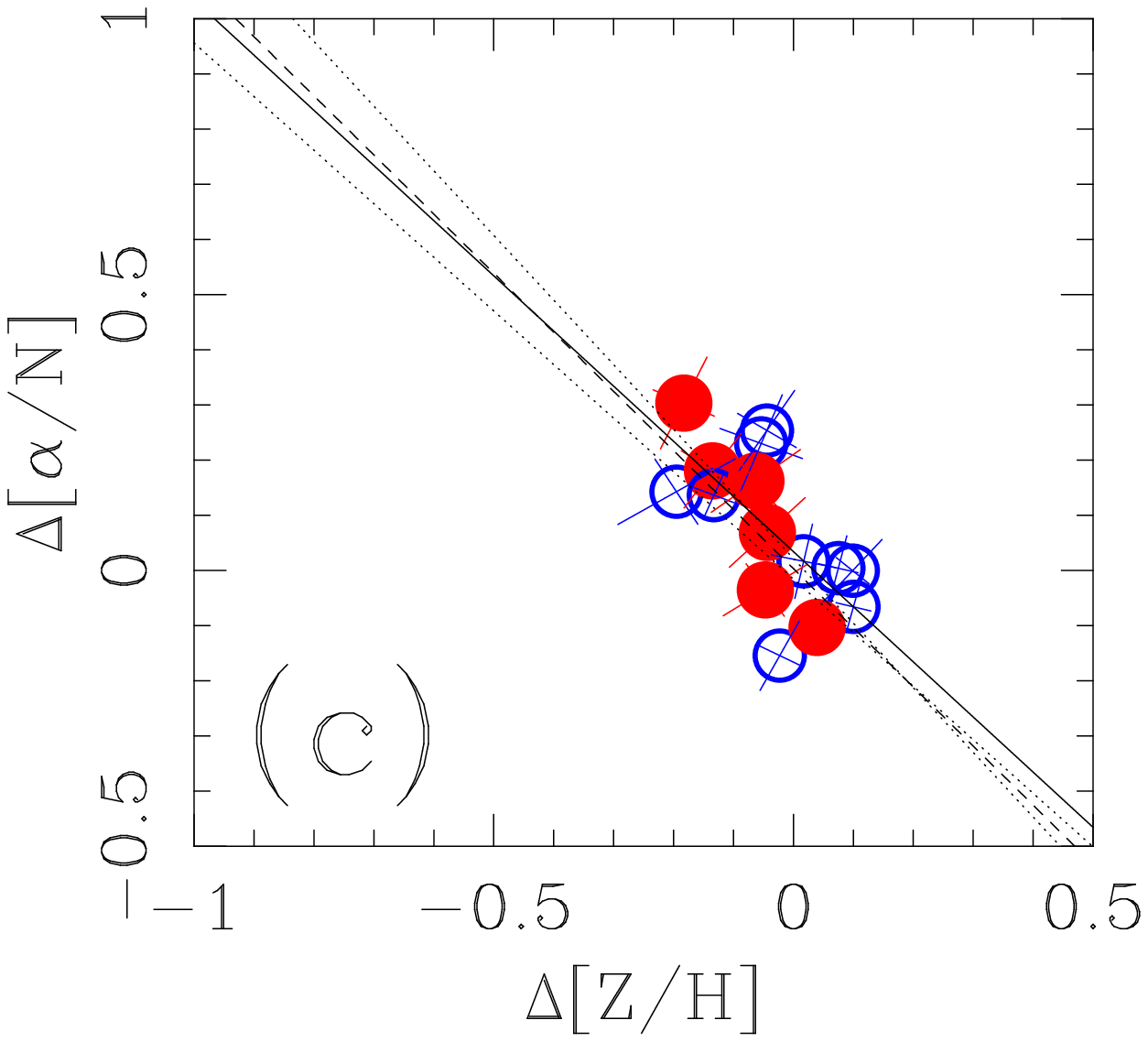}}
\hbox{\epsscale{0.25} \plotone{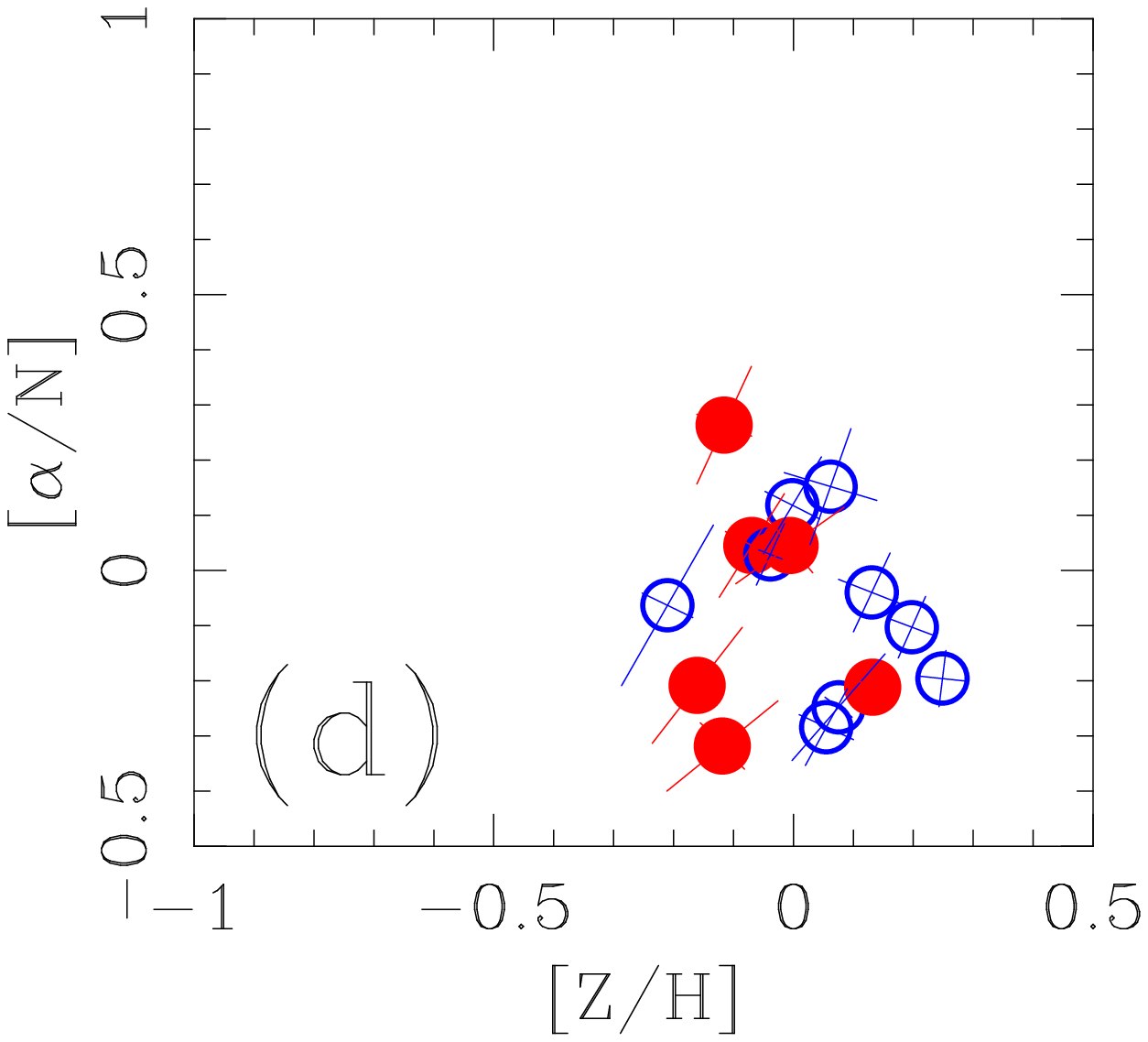}}
}
\caption{The relative and absolute age-metallicity and
nitrogen-metallicity diagrams for those E/S0 galaxies with $\sigma> 134$
km/s. (a) and (c) show the results from the differential fitting. (b)
and (d) show the parameters derived assuming no zero-point correction.
When the models are not recalibrated to the data, the results are quite
striking: (1) a ``statistically'' significant anti-correlation between
$\log t$ and [Z/H] appears, with a slope of $-0.74\pm 0.13$ dex/dex
(shown by the dashed and dotted lines); and (2) the nitrogen-metallicity
correlation completely disappears, masking the true nature of the
nitrogen enhancements. Systematic uncertainties in the calibrations of
our indices would have obscured the true nature of early-type stellar
populations. Note that the induced correlation between the absolute ages
and metallicities obeys the famed ``3/4'' rule \citep{worthey}.
\label{fig:abscor}}
\end{figure*}

In Figure \ref{fig:abscor} we show another comparison of the
differential and absolute analyses, in particular investigating the
effects on parameter-parameter correlations. For this discussion we
exclude those galaxies below $\sigma=134$ km/s to reduce the impact of
the luminosity-selection bias. In the left-hand panels we
show the relative ages and nitrogen enhancements plotted against
relative metallicity. No significant correlation exists between the
relative ages and relative metallicity. The significant correlation
between relative nitrogen enhancements and metallicity was discussed
earlier. If one ignores the recalibration of the models to our data, one
obtains an anti-correlation between age and metallicity with a slope of
$-0.76\pm 0.10$ dex/dex (shown by the dashed and dotted lines). Note
that this artificially induced slope obeys the famed ``3/4'' rule of
\cite{worthey} because the systematic errors in our line strengths
couple with the average age-metallicity degeneracy of the indices.

Furthermore, in the absolute analysis the correlation between nitrogen
enhancement and metallicity effectively disappears. Because Fig.
\ref{abs-early-feh-age}(b) shows the strong [Z/H]-$\sigma$ and
[N/$\alpha$]-$\sigma$ correlations, a lack of an observed correlation
between [Z/H] and [N/$\alpha$] would therefore have masked the true
nature of the [N/$\alpha$]-$\sigma$ correlation. Note that while the
general trends of the SSP parameters with velocity dispersion were
preserved when the recalibration was ignored, the systematic errors in
the calibration had resulted in larger errors in the SSP parameters.
These errors appeared in Figure \ref{abs-early-feh-age}(b) as
larger scatter and increased inhomogeneity. However, these added errors
can be both correlated and uncorrelated, possibly imposing apparent
(anti)correlations between age and metallicity, or erasing underlying
correlations between metallicity and abundance ratios.

For samples of galaxies at high redshift, it remains technical challenge
to ensure that absorption line strengths from the models and data have
no systematic offsets. Extreme care must be taken to eliminate all
sources of systematic error in order to utilize the models' absolute
scales of age and abundance. Fortunately, these large systematic sources
of error can be eliminated if one is only interested in the relative
properties of galaxies within an homogeneous sample.


\subsection{Uncertainties in the Adopted Model Zero-point}

In \S \ref{sec:recalib} we adopted a set of parameters describing the
mean stellar population of the most massive E/S0 galaxies in CL1358+62.
These parameters are reprinted as Model 0 in Table \ref{tab:var-zpts}.
Here we investigate some of the effects one obtains by varying the
parameters of this reference point. Table \ref{tab:var-zpts} lists the
standard parameters and the three additional variations, Models A, B, and
C. These variations are not meant as formal uncertainties in age,
[Z/H], or $\alpha$-enhancement, but were specifically chosen to be
significantly larger than the uncertainties in those parameters. As a
result, the general effects on our results may be considered very
conservative systematic uncertainties in our conclusions. Also note that
models in which the reference point has solar metallicity {\it and\/}
solar $\alpha$/Fe have been ruled out by many authors.

\begin{figure*}
\leftline{\epsscale{0.95}\plotone{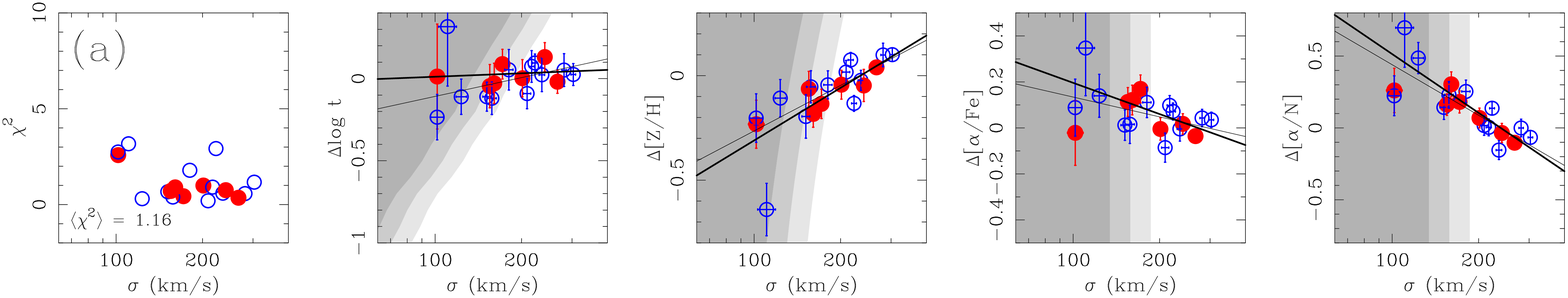}}
\leftline{\epsscale{0.95}\plotone{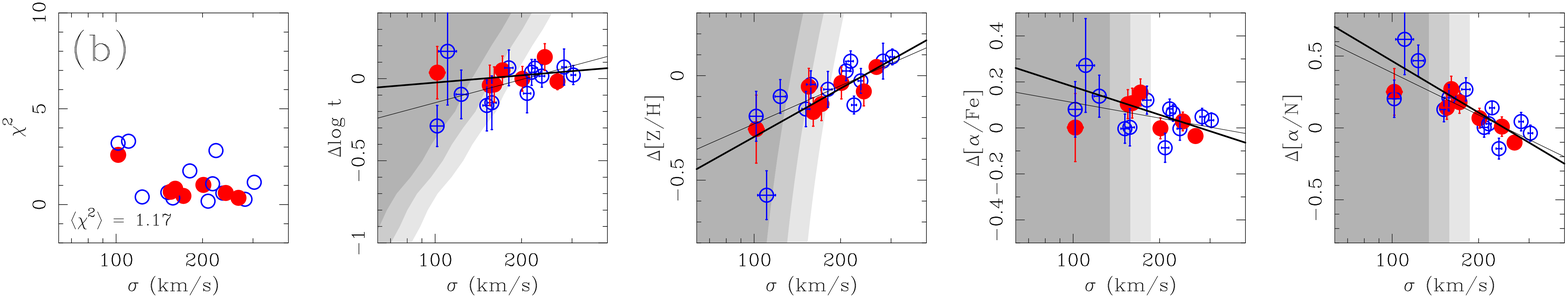}}
\leftline{\epsscale{0.95}\plotone{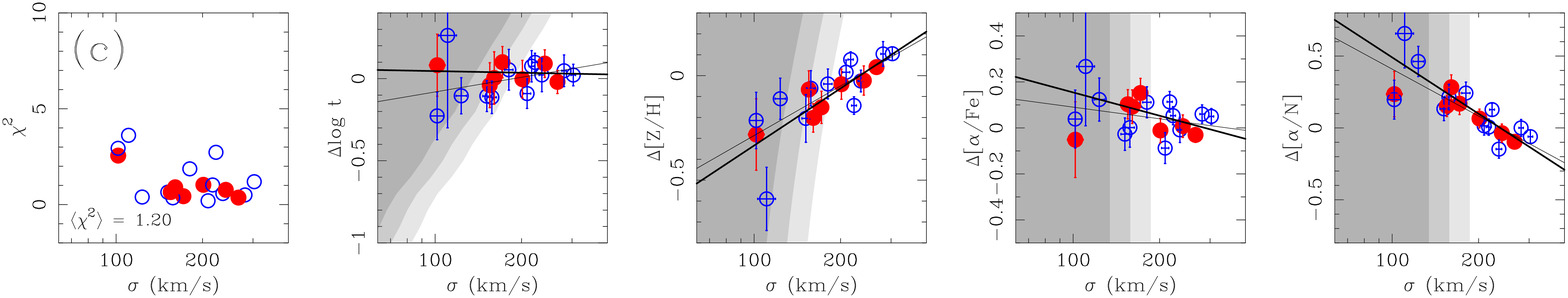}}
\leftline{\epsscale{0.95}\plotone{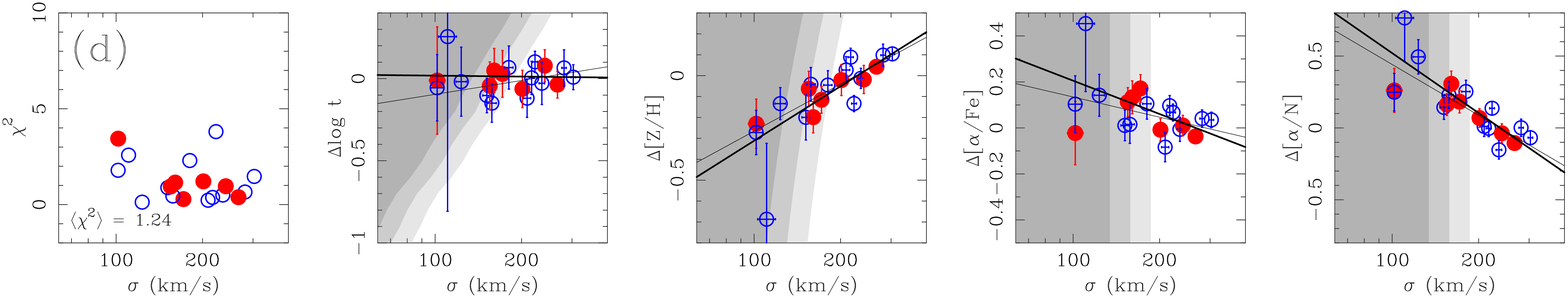}}
\caption{The effects of varying the model zero-point on the relative
stellar population parameters as functions of velocity dispersion. In
(a), (b), (c), and (d) we show the results
using the zero-points defined by Models 0, A, B, and C, respectively,
defined in Table \ref{tab:var-zpts}. Note that (a) is identical to Fig.
\ref{early-feh-age}(c). The general trends, and scatter about these
trends, are insensitive to these large changes in the adopted
zero-point. Because the variations in the recalibration probed here are
much larger than the true uncertainties in the reference parameters,
we conclude that our results are insensitive to systematic errors in the
recalibration.
\label{fig:rel-early-feh-age}}
\end{figure*}

Figure \ref{fig:rel-early-feh-age} shows the best-fit relative stellar
population parameters plotted against velocity dispersion, just as in
Figure \ref{early-feh-age}(c). No change in slope in any of the
correlations of SSP parameters with velocity dispersion is statistically
significant. Because the true uncertainties in the zero-point are
likely to be much smaller than the range probed by our variations in
age, metallicity, and $\alpha$-enhancement, we conclude that our results
are not sensitive to systematic uncertainties in the recalibration.

\clearpage

\LongTables

\begin{deluxetable}{llrrrrrr}
\tablecaption{Corrected Velocity Dispersions and Balmer Line
Strengths\label{tab:corindex-b}}
\tablewidth{0pt}
\tablehead{
\colhead{ID}&
\colhead{Type}&
\colhead{$\log\sigma$}&
\colhead{H$\delta_A$} &
\colhead{H$\delta_F$} &
\colhead{H$\gamma_A$} &
\colhead{H$\gamma_F$} &
\colhead{H$\beta$}}
\startdata
  095 &   S0&$  2.321\pm 0.013$ &$  -1.37\pm  0.57$ &$   0.22\pm  0.17$ &$  -5.11\pm  0.30$ &$  -1.50\pm  0.17$ &$   1.72\pm  0.13$ \\
  135 &   S0&$  2.180\pm 0.014$ &$  -0.35\pm  0.28$ &$   1.01\pm  0.18$ &$  -3.83\pm  0.27$ &$  -0.62\pm  0.16$ &$   2.43\pm  0.16$ \\
  182 &   S0&$  2.091\pm 0.018$ &$  -0.35\pm  0.40$ &$   0.38\pm  0.27$ &$  -3.49\pm  0.40$ &$  -0.63\pm  0.23$ &$   1.87\pm  0.27$ \\
  211 &   S0&$  2.257\pm 0.015$ &$  -0.97\pm  0.29$ &$   0.51\pm  0.19$ &$  -4.47\pm  0.31$ &$  -0.93\pm  0.19$ &$   2.21\pm  0.19$ \\
  212 &    E&$  2.235\pm 0.014$ &$  -0.52\pm  0.28$ &$   0.52\pm  0.20$ &$  -4.41\pm  0.23$ &$  -0.70\pm  0.13$ &$   1.79\pm  0.13$ \\
  215 &   S0&$  2.198\pm 0.015$ &$  -0.88\pm  0.38$ &$   0.50\pm  0.25$ &$  -4.41\pm  0.39$ &$  -1.09\pm  0.24$ &$   2.29\pm  0.21$ \\
  233 & E/S0&$  2.348\pm 0.008$ &$  -0.97\pm  0.17$ &$   0.92\pm  0.10$ &$  -4.60\pm  0.17$ &$  -0.88\pm  0.10$ &$   2.15\pm  0.09$ \\
  236 &   S0&$  2.271\pm 0.012$ & \nodata  & \nodata  &$  -5.10\pm  0.37$ &$  -0.83\pm  0.18$ &$   1.86\pm  0.21$ \\
  242 &    E&$  2.304\pm 0.011$ &$  -1.27\pm  0.22$ &$   0.51\pm  0.14$ &$  -4.97\pm  0.25$ &$  -1.44\pm  0.16$ &$   1.37\pm  0.14$ \\
  256 &    E&$  2.395\pm 0.007$ & \nodata  &$   0.43\pm  0.10$ &$  -4.85\pm  0.16$ &$  -1.14\pm  0.08$ &$   1.75\pm  0.08$ \\
  269 & E/S0&$  2.482\pm 0.008$ &$  -1.78\pm  0.16$ &$   0.21\pm  0.10$ &$  -5.45\pm  0.20$ &$  -1.37\pm  0.11$ &$   1.84\pm  0.08$ \\
  292 &   S0&$  2.028\pm 0.012$ & \nodata  & \nodata  &$  -2.21\pm  0.25$ &$   0.66\pm  0.16$ &$   2.38\pm  0.16$ \\
  298 &   S0&$  2.449\pm 0.008$ &$  -1.86\pm  0.19$ &$   0.51\pm  0.12$ &$  -5.29\pm  0.21$ &$  -1.39\pm  0.13$ &$   1.85\pm  0.09$ \\
  300 &   S0&$  2.372\pm 0.010$ &$  -1.29\pm  0.21$ &$   0.61\pm  0.15$ &$  -5.21\pm  0.26$ &$  -1.50\pm  0.15$ &$   1.58\pm  0.13$ \\
  303 &    E&$  2.191\pm 0.011$ &$  -0.55\pm  0.22$ &$   0.80\pm  0.13$ &$  -4.21\pm  0.26$ &$  -0.70\pm  0.15$ &$   1.94\pm  0.15$ \\
  309 & E/S0&$  2.336\pm 0.009$ &$  -1.31\pm  0.20$ &$   0.37\pm  0.13$ &$  -5.49\pm  0.24$ &$  -1.51\pm  0.14$ &$   1.69\pm  0.10$ \\
  353 & E/S0&$  2.336\pm 0.008$ & \nodata  & \nodata  &$  -5.18\pm  0.21$ &$  -1.41\pm  0.11$ &$   1.52\pm  0.09$ \\
  359 &   S0&$  2.279\pm 0.010$ & \nodata  & \nodata  &$  -4.09\pm  0.25$ &$  -0.73\pm  0.15$ &$   1.92\pm  0.12$ \\
  360 &    E&$  2.226\pm 0.048$ & \nodata  & \nodata  &$  -3.32\pm  0.97$ &$  -1.08\pm  0.63$ &$   1.52\pm  0.52$ \\
  375 &    E&$  2.470\pm 0.010$ & \nodata  & \nodata  &$  -5.69\pm  0.35$ &$  -1.92\pm  0.30$ &$   1.54\pm  0.58$ \\
  381 & E/S0&$  2.304\pm 0.011$ & \nodata  & \nodata  &$  -4.54\pm  0.26$ &$  -0.95\pm  0.15$ &$   1.83\pm  0.15$ \\
  391 & E/S0&$  2.392\pm 0.011$ & \nodata  &$   0.74\pm  0.23$ &$  -4.49\pm  0.21$ &$  -0.96\pm  0.13$ &$   1.83\pm  0.11$ \\
  408 &   S0&$  2.352\pm 0.011$ & \nodata  &$   0.80\pm  0.18$ &$  -4.77\pm  0.26$ &$  -1.10\pm  0.14$ &$   1.75\pm  0.11$ \\
  409 &    E&$  2.008\pm 0.027$ &$  -0.85\pm  0.51$ &$   0.69\pm  0.29$ &$  -4.39\pm  0.58$ &$  -0.62\pm  0.34$ &$   2.39\pm  0.41$ \\
  410 &   S0&$  2.149\pm 0.015$ & \nodata  & \nodata  &$  -4.71\pm  0.41$ &$  -0.99\pm  0.23$ &$   2.54\pm  0.23$ \\
  412 &    E&$  2.206\pm 0.016$ &$  -0.30\pm  0.27$ &$   0.76\pm  0.18$ &$  -3.85\pm  0.27$ &$  -0.56\pm  0.18$ &$   1.52\pm  0.20$ \\
  463 &   S0&$  2.431\pm 0.010$ & \nodata  & \nodata  &$  -4.52\pm  0.22$ &$  -0.97\pm  0.13$ &$   1.51\pm  0.14$ \\
  481 &   S0&$  2.007\pm 0.022$ &$  -0.04\pm  0.42$ &$   1.73\pm  0.25$ &$  -2.03\pm  0.48$ &$   0.14\pm  0.32$ &$   2.47\pm  0.31$ \\
  493 & E/S0&$  2.043\pm 0.031$ &$   1.85\pm  0.57$ &$   1.99\pm  0.38$ &$  -4.38\pm  0.60$ &$  -0.94\pm  0.36$ &$   2.03\pm  0.44$ \\
  531 &    E&$  2.426\pm 0.008$ &$  -1.68\pm  0.12$ &$   0.35\pm  0.09$ &$  -5.30\pm  0.17$ &$  -1.37\pm  0.09$ &$   1.77\pm  0.08$ \\
  534 &    E&$  2.060\pm 0.022$ & \nodata  & \nodata  &$  -3.51\pm  0.43$ &$  -0.35\pm  0.26$ &$   2.00\pm  0.26$ \\
  536 &    E&$  2.382\pm 0.009$ &$  -1.68\pm  0.20$ &$   0.56\pm  0.13$ &$  -5.28\pm  0.22$ &$  -1.45\pm  0.13$ &$   1.79\pm  0.10$ \\
\enddata
\end{deluxetable}

\clearpage
\begin{deluxetable}{ l  r r r r r r r r r r  }
\tablecaption{Corrected Molecular and Metal Line Strengths
\label{tab:corindex-m} }
\tablewidth{0pt}
\tablehead{
\colhead{ID}&
\colhead{CN1} &
\colhead{CN2} &
\colhead{Ca4227\AA} &
\colhead{G4300\AA} &
\colhead{Fe4383\AA} &
\colhead{Ca4455\AA} &
\colhead{Fe4531\AA} &
\colhead{C4668\AA} &
\colhead{Fe5015\AA} &
\colhead{Mg$b$}}
\startdata
  095 &$   0.06\pm  0.01$ &$   0.08\pm  0.01$ &$   0.69\pm  0.34$ &$   5.11\pm  0.26$ &$   4.47\pm  0.32$ &$   1.04\pm  0.19$ &$   3.24\pm  0.25$ &$   6.02\pm  0.35$ &$   4.83\pm  0.34$ &$   4.27\pm  0.29$ \\
  135 &$   0.01\pm  0.01$ &$   0.04\pm  0.01$ &$   0.75\pm  0.15$ &$   5.02\pm  0.23$ &$   3.73\pm  0.39$ &$   1.00\pm  0.22$ &$   2.97\pm  0.28$ &$   3.95\pm  0.76$ &$   5.32\pm  0.38$ & \nodata  \\
  182 &$  -0.03\pm  0.01$ &$   0.01\pm  0.01$ &$   1.09\pm  0.17$ &$   5.12\pm  0.36$ &$   2.87\pm  0.60$ &$   0.43\pm  0.34$ &$   2.81\pm  0.36$ &$   5.07\pm  0.57$ &$   2.88\pm  0.39$ & \nodata  \\
  211 &$   0.03\pm  0.01$ &$   0.06\pm  0.01$ &$   1.27\pm  0.12$ &$   5.35\pm  0.28$ &$   4.05\pm  0.44$ &$   1.12\pm  0.36$ &$   2.57\pm  0.27$ &$   5.29\pm  0.55$ &$   4.48\pm  0.41$ & \nodata  \\
  212 &$   0.03\pm  0.01$ &$   0.06\pm  0.01$ &$   0.91\pm  0.13$ &$   5.41\pm  0.21$ &$   2.98\pm  0.34$ &$   0.87\pm  0.20$ &$   2.65\pm  0.22$ &$   5.45\pm  0.53$ &$   4.29\pm  0.29$ &$   2.74\pm  0.38$ \\
  215 &$   0.02\pm  0.01$ &$   0.04\pm  0.01$ &$   0.78\pm  0.19$ &$   4.99\pm  0.33$ &$   3.30\pm  0.48$ &$   0.89\pm  0.27$ &$   3.07\pm  0.37$ &$   5.68\pm  0.47$ &$   4.38\pm  0.36$ &$   2.99\pm  0.43$ \\
  233 &$   0.03\pm  0.01$ &$   0.06\pm  0.01$ &$   0.87\pm  0.08$ &$   5.61\pm  0.13$ &$   3.88\pm  0.21$ &$   1.07\pm  0.11$ &$   3.34\pm  0.16$ &$   5.06\pm  0.26$ &$   4.58\pm  0.22$ & \nodata  \\
  236 &$   0.04\pm  0.01$ &$   0.07\pm  0.01$ &$   0.96\pm  0.17$ &$   5.43\pm  0.26$ &$   4.27\pm  0.55$ &$   1.05\pm  0.21$ &$   3.39\pm  0.25$ &$   6.14\pm  0.51$ &$   4.64\pm  0.31$ &$   4.20\pm  0.30$ \\
  242 &$   0.05\pm  0.01$ &$   0.08\pm  0.01$ &$   1.06\pm  0.11$ &$   5.05\pm  0.20$ &$   4.01\pm  0.23$ &$   1.18\pm  0.15$ &$   2.87\pm  0.17$ &$   5.79\pm  0.77$ &$   4.48\pm  0.28$ &$   3.99\pm  0.26$ \\
  256 &$   0.06\pm  0.01$ &$   0.09\pm  0.01$ &$   1.08\pm  0.11$ &$   5.35\pm  0.14$ &$   4.24\pm  0.36$ &$   1.19\pm  0.29$ &$   3.31\pm  0.14$ &$   6.11\pm  0.27$ &$   4.83\pm  0.17$ &$   3.96\pm  0.29$ \\
  269 &$   0.10\pm  0.01$ &$   0.13\pm  0.01$ &$   1.09\pm  0.09$ &$   5.49\pm  0.14$ &$   4.43\pm  0.24$ &$   1.17\pm  0.11$ &$   3.00\pm  0.19$ &$   7.16\pm  0.24$ &$   4.85\pm  0.19$ &$   4.31\pm  0.24$ \\
  292 & \nodata  & \nodata  &$   0.76\pm  0.14$ &$   4.55\pm  0.21$ &$   3.38\pm  0.36$ &$   0.70\pm  0.27$ &$   2.86\pm  0.21$ &$   4.52\pm  0.39$ &$   4.67\pm  0.33$ &$   2.02\pm  0.28$ \\
  298 &$   0.08\pm  0.01$ &$   0.12\pm  0.01$ &$   1.24\pm  0.12$ &$   5.51\pm  0.16$ &$   4.33\pm  0.29$ &$   1.21\pm  0.12$ &$   3.41\pm  0.17$ &$   6.97\pm  0.81$ &$   5.45\pm  0.20$ &$   4.18\pm  0.40$ \\
  300 &$   0.07\pm  0.01$ &$   0.12\pm  0.01$ &$   0.92\pm  0.11$ &$   5.30\pm  0.21$ &$   4.34\pm  0.35$ &$   1.43\pm  0.37$ &$   3.15\pm  0.19$ &$   5.75\pm  0.36$ &$   5.03\pm  0.29$ &$   4.20\pm  0.27$ \\
  303 &$   0.04\pm  0.01$ &$   0.06\pm  0.01$ &$   0.92\pm  0.20$ &$   5.42\pm  0.20$ &$   3.90\pm  0.62$ &$   1.07\pm  0.32$ &$   3.12\pm  0.21$ &$   5.15\pm  0.70$ &$   4.80\pm  0.27$ &$   3.32\pm  0.37$ \\
  309 &$   0.08\pm  0.01$ &$   0.12\pm  0.01$ &$   0.97\pm  0.12$ &$   5.61\pm  0.19$ &$   3.75\pm  0.32$ &$   0.75\pm  0.25$ &$   3.27\pm  0.18$ &$   7.00\pm  0.32$ &$   5.04\pm  0.24$ &$   4.00\pm  0.27$ \\
  353 & \nodata  & \nodata  &$   0.75\pm  0.12$ &$   5.25\pm  0.16$ &$   3.95\pm  0.26$ &$   0.88\pm  0.16$ &$   3.06\pm  0.16$ &$   6.04\pm  0.31$ &$   4.43\pm  0.21$ &$   3.78\pm  0.20$ \\
  359 & \nodata  & \nodata  &$   0.85\pm  0.14$ &$   4.79\pm  0.21$ &$   3.73\pm  0.46$ &$   1.05\pm  0.32$ &$   3.25\pm  0.20$ &$   5.36\pm  0.39$ &$   3.84\pm  0.33$ &$   3.40\pm  0.32$ \\
  360 & \nodata  & \nodata  &$  -0.15\pm  0.66$ &$   4.10\pm  0.86$ &$   2.95\pm  1.61$ &$   1.13\pm  0.92$ &$   2.40\pm  0.75$ &$   2.44\pm  1.20$ &$   8.18\pm  0.89$ &$   4.36\pm  0.72$ \\
  375 &$   0.12\pm  0.01$ &$   0.15\pm  0.01$ &$   0.95\pm  0.10$ &$   5.40\pm  0.15$ &$   4.37\pm  0.26$ &$   1.19\pm  0.09$ &$   3.48\pm  0.14$ &$   6.82\pm  0.27$ &$   5.50\pm  0.41$ &$   4.45\pm  0.37$ \\
  381 &$   0.04\pm  0.03$ &$   0.07\pm  0.02$ &$   0.91\pm  0.12$ &$   5.15\pm  0.27$ &$   4.02\pm  0.29$ &$   0.99\pm  0.18$ &$   2.98\pm  0.22$ &$   5.00\pm  0.50$ &$   4.31\pm  0.34$ &$   3.87\pm  0.29$ \\
  391 &$   0.05\pm  0.01$ &$   0.08\pm  0.01$ &$   1.22\pm  0.14$ &$   5.07\pm  0.18$ &$   4.00\pm  0.32$ &$   1.02\pm  0.24$ &$   3.15\pm  0.19$ &$   5.41\pm  0.48$ &$   4.50\pm  0.36$ &$   3.54\pm  0.43$ \\
  408 &$   0.03\pm  0.01$ &$   0.06\pm  0.01$ &$   0.46\pm  0.33$ &$   5.66\pm  0.23$ &$   3.94\pm  0.32$ &$   1.10\pm  0.28$ &$   3.19\pm  0.20$ &$   6.20\pm  0.34$ &$   4.18\pm  0.33$ &$   3.66\pm  0.39$ \\
  409 &$  -0.01\pm  0.01$ &$   0.02\pm  0.02$ &$   0.76\pm  0.32$ &$   4.88\pm  0.53$ &$   2.55\pm  0.68$ &$   0.71\pm  0.40$ &$   4.19\pm  0.46$ &$   4.07\pm  0.92$ &$   3.57\pm  0.75$ &$   4.24\pm  0.70$ \\
  410 & \nodata  & \nodata  &$   1.08\pm  0.21$ &$   5.18\pm  0.36$ &$   3.62\pm  0.48$ &$   1.21\pm  0.29$ &$   3.93\pm  0.34$ &$   6.75\pm  0.52$ &$   4.35\pm  0.41$ &$   4.28\pm  0.75$ \\
  412 &$   0.01\pm  0.01$ &$   0.03\pm  0.01$ &$   1.19\pm  0.40$ &$   5.32\pm  0.24$ &$   3.30\pm  0.32$ &$   0.95\pm  0.27$ &$   2.57\pm  0.23$ &$   4.42\pm  0.52$ &$   4.38\pm  0.39$ &$   3.35\pm  0.24$ \\
  463 & \nodata  & \nodata  &$   1.02\pm  0.11$ &$   5.10\pm  0.18$ &$   4.16\pm  0.28$ &$   1.09\pm  0.17$ &$   3.31\pm  0.19$ &$   6.20\pm  0.75$ &$   4.46\pm  0.24$ &$   3.99\pm  0.30$ \\
  481 &$  -0.01\pm  0.01$ &$   0.03\pm  0.01$ &$   1.23\pm  0.23$ &$   4.49\pm  0.42$ &$   2.53\pm  0.71$ &$   1.14\pm  0.39$ &$   2.87\pm  0.42$ &$   3.66\pm  0.75$ &$   4.64\pm  0.60$ &$   3.47\pm  0.45$ \\
  493 &$  -0.07\pm  0.02$ &$  -0.04\pm  0.02$ &$   0.44\pm  0.34$ &$   5.41\pm  0.56$ &$   2.28\pm  0.94$ &$   1.39\pm  0.55$ &$   3.06\pm  0.59$ &$   0.88\pm  1.14$ &$   3.99\pm  0.74$ &$   3.59\pm  1.07$ \\
  531 &$   0.08\pm  0.00$ &$   0.12\pm  0.00$ &$   0.99\pm  0.09$ &$   5.33\pm  0.14$ &$   4.54\pm  0.18$ &$   1.16\pm  0.09$ &$   3.15\pm  0.14$ &$   6.35\pm  0.22$ &$   4.63\pm  0.20$ &$   4.37\pm  0.23$ \\
  534 & \nodata  & \nodata  & \nodata  &$   4.15\pm  0.35$ &$   3.07\pm  1.72$ & \nodata  &$   2.60\pm  0.38$ &$   3.80\pm  0.92$ &$   1.93\pm  0.85$ & \nodata  \\
  536 &$   0.07\pm  0.01$ &$   0.10\pm  0.01$ &$   1.11\pm  0.11$ &$   5.30\pm  0.17$ &$   3.99\pm  0.29$ &$   1.16\pm  0.13$ &$   3.06\pm  0.16$ &$   6.04\pm  0.77$ &$   4.74\pm  0.23$ &$   3.88\pm  0.28$ \\
\enddata
\end{deluxetable}

\clearpage

\begin{deluxetable}{l r r r r r r}
\tabletypesize{\small}
\tablewidth{0pt}
\tablecaption{First Partial Derivatives of the Models
\label{tab:derivs}
}
\tablehead{
\colhead{$X$} &
\colhead{$\partial X\over\partial \log t$} &
\colhead{$\partial X\over\partial [{\rm Z/H}]$} &
\colhead{$\partial X\over\partial [\alpha{\rm /Fe}]$} &
\colhead{$\partial X\over\partial [\alpha/{\rm N}]$} &
\colhead{$\partial X\over\partial [\alpha/{\rm C}]$} &
\colhead{$\partial X\over\partial [\alpha/{\rm Ca}]$}
}
\startdata
  H$\delta_F$ & $ -1.813$ & $ -1.356$ & $  2.064$ & $ \ldots$ & $ \ldots$ & $ \ldots$  \\
  H$\delta_A$ & $ -4.299$ & $ -4.185$ & $  4.743$ & $ \ldots$ & $ \ldots$ & $ \ldots$  \\
       CN$_1$ & $  0.122$ & $  0.198$ & $  0.058$ & $ -0.147$ & $ -0.474$ & $ \ldots$  \\
       CN$_2$ & $  0.126$ & $  0.213$ & $  0.066$ & $ -0.164$ & $ -0.507$ & $ \ldots$  \\
       Ca4227 & $  1.000$ & $  1.444$ & $ -0.034$ & $  0.421$ & $  2.259$ & $ -1.767$  \\
        G4300 & $  1.462$ & $  1.748$ & $  1.584$ & $ \ldots$ & $ -4.779$ & $ \ldots$  \\
  H$\gamma_F$ & $ -2.612$ & $ -2.232$ & $  1.181$ & $ \ldots$ & $ \ldots$ & $ \ldots$  \\
  H$\gamma_A$ & $ -4.229$ & $ -4.637$ & $  3.788$ & $ \ldots$ & $ \ldots$ & $ \ldots$  \\
       Fe4383 & $  2.156$ & $  4.920$ & $ -4.488$ & $ \ldots$ & $ \ldots$ & $ \ldots$  \\
       Ca4455 & $  0.693$ & $  1.434$ & $  0.104$ & $ \ldots$ & $ \ldots$ & $ \ldots$  \\
       Fe4531 & $  0.973$ & $  2.084$ & $ -1.013$ & $ \ldots$ & $ \ldots$ & $ \ldots$  \\
        C4668 & $  2.142$ & $  8.953$ & $  1.209$ & $ \ldots$ & $-50.216$ & $ \ldots$  \\
     H$\beta$ & $ -1.154$ & $ -0.629$ & $  0.289$ & $ \ldots$ & $ \ldots$ & $ \ldots$  \\
       Fe5015 & $  1.001$ & $  3.208$ & $ -1.574$ & $ \ldots$ & $ \ldots$ & $ \ldots$  \\
        Mg$b$ & $  1.761$ & $  3.183$ & $  2.123$ & $ \ldots$ & $ \ldots$ & $ \ldots$  \\
\enddata
\tablecomments{
The partial derivatives given above are computed by {\it adding\/} 0.3
dex to the adopted reference parameters $ \{\log t, [{\rm Z/H}] ,
[\alpha{\rm /Fe}], [\alpha/{\rm N}], [\alpha/{\rm C}], [\alpha/{\rm
Ca}]\} = \{ -0.15, 0.3, 0.2, 0, 0, 0\}$. Note that we define age in
units of $10^{10}$ yr, and thus the reference age is equivalent to $t=7$
Gyr, or a redshift of formation of $z_f=2.4$ \citep{kelson01}.
Thus $\partial X/\partial p \equiv \Delta X/0.3$, where
$ p \in \{\log t, [{\rm Z/H}] , [\alpha{\rm /Fe}], [\alpha/{\rm N}],
[\alpha/{\rm C}], [\alpha/{\rm Ca}]\}$.
Where the partial derivatives are not given, and assumed to be
identically zero, \cite{thomas} and \cite{thomas2} had {\it either}
reported, qualitatively, that the index was insensitive to changes in the
relevant parameter, {\it or} the model cannot contain information on the
sensitivity of the index to the parameter because they did not (or could
not) include it in their analyses. This fact helps to underscore the
importance of zero-pointing the model to a fiducial observed dataset as
it removes the importance of such unknown/undefined dependences to
zeroth order. The \cite{thomas} and \cite{thomas2} set of predictions
is highly non-linear such that at other locations in the 6D parameter
domain these derivatives are not valid. As a result, the above partial
derivatives should only illustrate a broad sensitivity of the indices to
changes in the model parameters.
}
\end{deluxetable}

\begin{deluxetable}{l c c c l}
\tablewidth{0pt}
\tablecaption{The Zero-point Offsets for Recalibrating the Models
to the Data
\label{tab:zpts}
}
\tablehead{
\colhead{$X$} &
\colhead{$X_{\rm model}$} &
\colhead{$X_{\rm \langle CL1358\rangle}$} &
\colhead{$\Delta_X$} &
\colhead{units}
}
\startdata
  H$\delta_F$ & $  0.261$ & $  0.519$ & $  0.258$ & \AA \\
  H$\delta_A$ & $ -2.109$ & $ -1.468$ & $  0.641$ & \AA \\
       CN$_1$ & $  0.080$ & $  0.060$ & $ -0.020$ & mag \\
       CN$_2$ & $  0.113$ & $  0.094$ & $ -0.020$ & mag \\
       Ca4227 & $  1.432$ & $  0.958$ & $ -0.474$ & \AA \\
        G4300 & $  6.042$ & $  5.326$ & $ -0.716$ & \AA \\
  H$\gamma_F$ & $ -1.461$ & $ -1.265$ & $  0.196$ & \AA \\
  H$\gamma_A$ & $ -5.829$ & $ -5.004$ & $  0.825$ & \AA \\
       Fe4383 & $  5.496$ & $  4.138$ & $ -1.358$ & \AA \\
       Ca4455 & $  1.945$ & $  1.096$ & $ -0.848$ & \AA \\
       Fe4531 & $  3.573$ & $  3.166$ & $ -0.407$ & \AA \\
        C4668 & $  7.228$ & $  6.074$ & $ -1.154$ & \AA \\
     H$\beta$ & $  1.755$ & $  1.729$ & $ -0.025$ & \AA \\
       Fe5015 & $  5.855$ & $  4.690$ & $ -1.165$ & \AA \\
        Mg$b$ & $  4.422$ & $  4.000$ & $ -0.422$ & \AA \\
\enddata
\tablecomments{
The model values given above are computed at
the adopted reference parameters $ \{\log t, [{\rm Z/H}] ,
[\alpha{\rm /Fe}], [\alpha/{\rm N}], [\alpha/{\rm C}], [\alpha/{\rm
Ca}]\} = \{ -0.15, 0.3, 0.2, 0, 0, 0\}$. Note that we define age in
units of $10^{10}$ yr, and thus the reference age is equivalent to $t=7$
Gyr, or a redshift of formation of $z_f=2.4$ \citep{kelson01}.
}
\end{deluxetable}

\begin{deluxetable}{l c c c c c c}
\tabletypesize{\small}
\tablewidth{0pt}
\tablecaption{Tested Variations in the Zero-point of the Models
\label{tab:var-zpts}
}
\tablehead{
\colhead{Model} &
\colhead{$\log [t/(10 \hbox{ Gyr})]$} &
\colhead{[Z/H]} &
\colhead{[$\alpha$/Fe]} &
\colhead{[$\alpha$/N]} &
\colhead{[$\alpha$/C]} &
\colhead{[$\alpha$/Ca]}
}
\startdata
0$^a$    & $  -0.15$ & $ 0.30$ & $ 0.20$ & $ 0.00$ & $ 0.00$ & $ 0.00$\\
A        & $\,\,\ 0.00$ & $ 0.30$ & $ 0.30$ & $ 0.00$ & $ 0.00$ & $ 0.00$\\
B        & $  -0.15$ & $ 0.30$ & $ 0.00$ & $ 0.00$ & $ 0.00$ & $ 0.00$\\
C        & $  -0.15$ & $ 0.00$ & $ 0.00$ & $ 0.00$ & $ 0.00$ & $ 0.00$\\
\enddata
\tablecomments{
$^a$The zero-point used in the analysis presented in the paper.
}
\end{deluxetable}

\end{document}